\documentclass[amsmath,amssymb,
 aps,
prb,
twocolumn,
superscriptaddress,
]{revtex4-2}
\usepackage{hyperref}
\hypersetup{
colorlinks=true,
citecolor=blue,
linkcolor=blue,
urlcolor=blue,
}
\usepackage{xcolor}	
\usepackage{siunitx}
\usepackage{amsmath}
\usepackage{mathrsfs}
\usepackage{graphicx}
\usepackage{dcolumn}
\usepackage{bm}
\usepackage[normalem]{ulem}
\usepackage{float}
\usepackage{braket}
\usepackage{amssymb}
\usepackage{soul,xcolor}

\newcommand{\beq}{\begin{eqnarray}}
\newcommand{\eeq}{\end{eqnarray}}

\begin{document}

\title{Molecular spin qudits to test generalized Bell inequalities}

\author{Silvia Macedonio}
\affiliation{Dipartimento di Scienze Matematiche, Fisiche e Informatiche, Universit\`a  di Parma, Parco Area delle Scienze, 53/A, I-43124 Parma, Italy.}
\affiliation{Gruppo Collegato di Parma, INFN-Sezione Milano-Bicocca, I-43124 Parma, Italy.}

\author{Luca Lepori}
\affiliation{Dipartimento di Scienze Matematiche, Fisiche e Informatiche, Universit\`a  di Parma, Parco Area delle Scienze, 53/A, I-43124 Parma, Italy.}
\affiliation{Gruppo Collegato di Parma, INFN-Sezione Milano-Bicocca, I-43124 Parma, Italy.}

\author{Alessandro Chiesa}
\affiliation{Dipartimento di Scienze Matematiche, Fisiche e Informatiche, Universit\`a  di Parma, Parco Area delle Scienze, 53/A, I-43124 Parma, Italy.}
\affiliation{Gruppo Collegato di Parma, INFN-Sezione Milano-Bicocca, I-43124 Parma, Italy.}
\affiliation{UdR Parma, INSTM, I-43124 Parma, Italy.}

\author{Simone Chicco}
\affiliation{Dipartimento di Scienze Matematiche, Fisiche e Informatiche, Universit\`a  di Parma, Parco Area delle Scienze, 53/A, I-43124 Parma, Italy.}

\author{Laura Bersani}
\affiliation{Dipartimento di Scienze Matematiche, Fisiche e Informatiche, Universit\`a  di Parma, Parco Area delle Scienze, 53/A, I-43124 Parma, Italy.}

\author{Marcos Rubin-Osanz}
\affiliation{Dipartimento di Scienze Matematiche, Fisiche e Informatiche, Universit\`a  di Parma, Parco Area delle Scienze, 53/A, I-43124 Parma, Italy.}

\author{Lukas Bradley Woodcock}
\affiliation{Department of Chemistry, University of Copenhagen, DK-2100 Copenhagen, Denmark.}

\author{Athanasios Mavromagoulos}
\affiliation{Department of Chemistry, University of Copenhagen, DK-2100 Copenhagen, Denmark.}

\author{Giuseppe Allodi}
\affiliation{Dipartimento di Scienze Matematiche, Fisiche e Informatiche, Universit\`a  di Parma, Parco Area delle Scienze, 53/A, I-43124 Parma, Italy.}

\author{Elena Garlatti}
\affiliation{Dipartimento di Scienze Matematiche, Fisiche e Informatiche, Universit\`a  di Parma, Parco Area delle Scienze, 53/A, I-43124 Parma, Italy.}
\affiliation{Gruppo Collegato di Parma, INFN-Sezione Milano-Bicocca, I-43124 Parma, Italy.}
\affiliation{UdR Parma, INSTM, I-43124 Parma, Italy.}

\author{Stergios Piligkos}
\affiliation{Department of Chemistry, University of Copenhagen, DK-2100 Copenhagen, Denmark.}

\author{Augusto Smerzi}
\affiliation{QSTAR and INO-CNR and LENS, Largo Enrico Fermi 2, 50125 Firenze, Italy.}

\author{Stefano Carretta}
\email{stefano.carretta@unipr.it}
\affiliation{Dipartimento di Scienze Matematiche, Fisiche e Informatiche, Universit\`a  di Parma, Parco Area delle Scienze, 53/A, I-43124 Parma, Italy.}
\affiliation{Gruppo Collegato di Parma, INFN-Sezione Milano-Bicocca, I-43124 Parma, Italy.}
\affiliation{UdR Parma, INSTM, I-43124 Parma, Italy.}

\begin{abstract} 
We show that Yb(trensal) molecular nanomagnet, embedding an electronic spin qubit coupled to a nuclear spin qudit,  provides an ideal platform to probe entanglement in a qubit-qudit system. 
This is demonstrated by developing an optimized pulse sequence to show violation of generalized Bell inequalities and by performing 
realistic numerical simulations including experimentally measured decoherence. We find that the inequalities are safely violated in a wide 
range of parameters, proving the robustness of entanglement in the investigated system. Furthermore, we propose a scheme to study qudit-qudit entanglement on a molecular spin trimer, in which two spins  3/2 are linked via an interposed switch to turn on and off their mutual interaction.
\end{abstract} 

\maketitle

Entanglement is a cornerstone of quantum technologies, enabling critical advancements in quantum computing, cryptography and sensing by exploiting quantum correlations for tasks that are impossible with classical systems \cite{Horodecki_2009, Pezz__2018,gentini2024}.
The characterization of these correlations  is an increasing relevant problem for both foundational investigations and the advancement of quantum technologies. This is usually done by preparing an entangled state of two objects and then measuring local observables on the two subsystems. The results are then combined to test the so-called {\it Bell inequalities}, whose violation represents a signature of quantum correlations and is generally used as a benchmark for distinguishing quantum predictions from local hidden-variable (HV) models \cite{nielsen}.

So far, most experimental and theoretical efforts to verify Bell inequalities have focused on qubit systems  \cite{Dada_2011,Lo_2016}. 
The extension to qudits remains mostly unexplored, despite the
growing availability of them in platforms like molecular nanomagnets, trapped ions, Rydberg atoms, transmons \cite{tweezer1,tweezer2,tweezer3,chiesa2024,PhysRevApplied.23.034046}, as well as the increasing interest for qudit-based encodings for  quantum simulation, sensing and computation.
Indeed, unlike traditional qubits, hosting only two states, qudits have the capability to encode quantum information across multiple states, providing a lager Hilbert space per physical unit that can enhance computational power, increase information density, and improve resistance to errors \cite{
quditsandhighdim, Imany_2019}. For instance, they allow more compact encodings with reduced number of entangling gates for quantum computation \cite{
lanyon} or quantum simulation \cite{tacchino,roca,meth,ciavarella} and can embed quantum error correction within single objects \cite{newclass,albert,gross,mezzadri2024}, greatly simplifying its actual implementation.  
Moreover, high-dimensional systems have shown significant advantages in quantum key distribution and quantum communication, where they can enhance security, robustness against noise, and channel capacity \cite{Cozzolino_2019, haoyu}.

From a foundational perspective, qudit entanglement is not merely a higher-dimensional analog of qubit entanglement: it supports qualitatively different forms of nonclassical correlations. In this context, generalized Bell inequalities
for $d$-dimensional systems serve as useful tools for validating genuine high-dimensional entanglement \cite{Weiss_2016, fonseca, Dada_2011}. Moreover, similar to the qubit case, they are sufficient to verify the Kocken-Speker contextuality of hidden-variable models of quantum mechanics \cite{fine1982}. This property is considered a crucial resource for quantum computation advantages \cite{howard2014,karanjai2018}. Probing generalized Bell inequalities on qudits is therefore important for both fundamental and applicative issues. However, finding a suitable platform for that task is not trivial, especially due to the intrinsic difficulty in the coherent manipulation of multi-level systems.

A convenient experimental platform providing qudits with even large spin-manifold dimensions is offered by Molecular Nanomagnets (MNMs) \cite{chiesa2024}.
These solid-state systems have gained considerable attention as building blocks for qudit-based quantum information processing due to their unique characteristics. In particular, the long coherence times \cite{Zadrozny} and versatile spin dynamics of MNMs enable precise control and manipulation of these qudits \cite{sessoliatzori,atzoritesi}. Moreover, microscopic interactions, level structure, eigenstates and coherence properties of MNMs can often be tuned at the synthetic level to a large extent \cite{moreno,Switch,switch4,bennett,lockyer}.
This allows to leverage the multi-level structure of MNMs for robust error correction schemes and efficient implementation of quantum algorithms, e.g. for quantum simulation and sensing \cite{review,qec,mezzadri_2024bis, Carretta_2021, mezzadri2024,chiccoallodi,roca,tacchino}.

Here we propose the use of an existing Yb(trensal) Molecular Nanomagnet (MNM) to test generalized Bell inequalities on 
spin qudits. This system provides an effective electronic spin 1/2 coupled by a strong hyperfine interaction to a nuclear spin 5/2, showing remarkably long coherence times. Such a qubit-qudit pair represents the simplest setup to test Bell inequalities beyond the qubit-qubit case.  
In particular, we focus on four qudit levels and we propose a semi-device independent protocol for the experimental verification of the Clauser-Horne-Shimony-Holt (CHSH) generalized-Bell inequality on this  $\Big(\frac{1}{2} -\frac{3}{2}\Big)$ molecular spin dimer. 
Indeed, this inequality allows to detect entanglement
without trusting every detail of the measurement setup, but only with a few basic assumptions (like dimension and measurement settings) \cite{brunner2011,brunner2014}.

We design the sequence of pulses to generate entangled states and then to perform local unitary transformations before measuring expectation values on the two spins. Experiments to probe the considered inequalities are numerically simulated in  presence of the experimentally measured dephasing. We exploit the gradient ascent pulse engineering (GRAPE) algorithm \cite{lambert2024qutip5quantumtoolbox} to optimize the sequence and thus reduce the leakage and/or the pulse duration, thus greatly mitigating the effect of decoherence. As a result, we obtain a very high fidelity and we demonstrate the violation of the CHSH inequalities using experimentally measured  Hamiltonian parameters and coherence times and experimentally feasible pulses.  \\
In addition, we extend our protocol to  two entangled qudits, such as a molecular spin $\Big(\frac{3}{2} - \frac{3}{2}\Big)$ dimer. In this case, we evaluate the Collins-Gisin-Linden-Massar-Popescu (CGLMP) inequality \cite{Collins_2002}, explicitly designed for pairs of $d-$dimensional systems \cite{CHSH}. \\
Demonstrating Bell-type correlations in such systems, under realistic conditions and including the effects of decoherence, provides both a meaningful test of quantum mechanics and a benchmark for future applications in quantum simulation, metrology, semi-device independent certification and error-corrected computation.

{\it CHSH inequality for a $\big(\frac{1}{2} - \frac{3}{2}\big)$ system--}
The general formalism for CHSH-type inequalities on a qubit–qudit pair has been recently developed in Ref.~\cite{moreno2024}, generalizing the results obtained for qubit–qubit systems~\cite{3hor}. Indeed, it was previously shown \cite{pironio2014} that dichotomic measurements 
at the basis of the CHSH formalism are sufficient to probe the validity of hidden-variables statistical models also on qubit-qudit states. The net effect is that CHSH inequalities hold, if properly defined, also for the qubit-qudit states. 
In this context, each subsystem is assigned to a party—commonly referred to as “Alice” and “Bob”—who can perform one of two measurements on the respective subsystem. We denote by \( A \) and \( A' \) the two observables measured by Alice (on the qubit), and by \( B \) and \( B' \) the two observables measured by Bob (on the qudit). These measurements are repeated on multiple copies of the system to estimate the probabilities of the various joint outcomes~\cite{Bell,CHSH}.

\begin{figure*}[!ht]
    \centering
\includegraphics[width=1\textwidth]{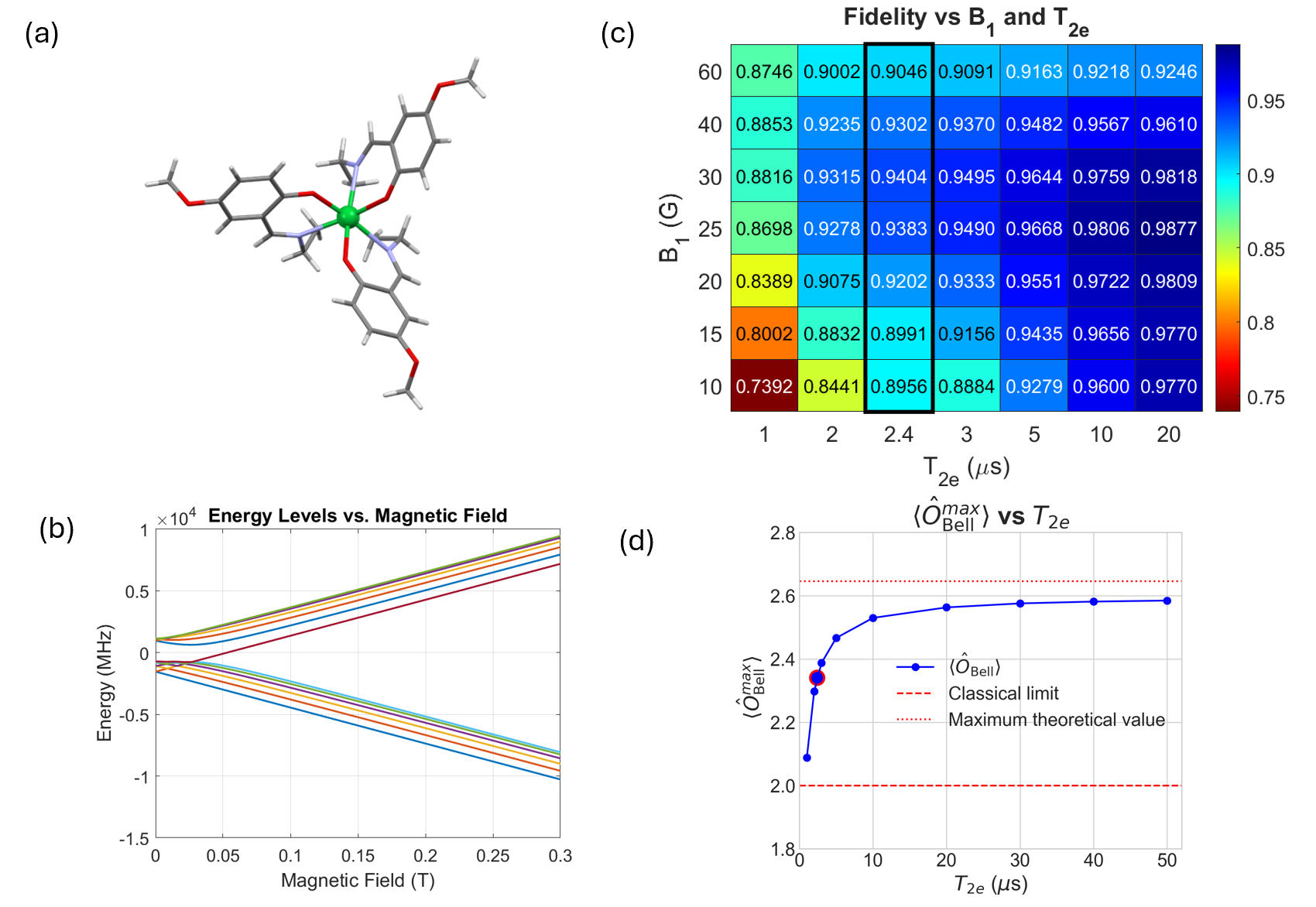}
    \caption{
        (a) Energy level diagram of the electro-nuclear dimer provided by the Yb(trensal) molecule (shown in the inset) as a function of the static magnetic field $B_0$ along $z$ axis. The system provides an $S=1/2-I=5/2$ spin pair [sketched in panel (b)] coupled by hyperfine interaction, used to probe CHSH inequality for $B_0 = 0.3$ T.
        (c) Fidelity $\mathcal{F}$ as a function of control field amplitude $B_1$ and decoherence time $T_{2e}$ in the preparation of the entangled state $\ket{\psi}$. The red box marks the experimentally measured $T_{2e}$.
        (d) $\langle \hat{O}^{(\mathrm{max})}_{\mathrm{Bell}} \rangle$ as a function of decoherence time $T_{2e}$, starting from the optimal entangled states. Violation of the CHSH inequality is obtained for $\langle \hat{O}^{(\mathrm{max})}_{\mathrm{Bell}} \rangle>2$ (red dashed line).}
    \label{figura_completa}
\end{figure*}

To test local hidden-variable (HV) models, one considers the expectation value of the Bell operator
\begin{equation}
\hat{O}_{\rm Bell} = A \otimes (B + B') + A' \otimes (B - B') \, ,
\label{bellexp}
\end{equation}
where \( A, A' \) and \( B, B' \) are Hermitian observables with eigenvalues \( \{+1, -1\} \) (see Ref.~\cite{moreno2024} and Supplementary Material). A violation of the CHSH inequality, excluding the possibility of a local HV description, occurs whenever
\begin{equation}
\langle \psi | \hat{O}_{\rm Bell} | \psi \rangle > 2 \, ,
\label{ineq}
\end{equation}
for a composite quantum state \( \ket{\psi} \).
Such a violation reveals the presence of nonclassical correlations and hence  the genuinely quantum nature of the joint qubit-qudit state $\ket{\psi}$. We focus hereafter on the following entangled qubit-qudit state for a spin \((\frac{1}{2}-\frac{3}{2})\) dimer
\beq
\ket{\psi} = \frac{1}{2} \, \Big(\ket{\uparrow,\frac{3}{2}} + \ket{\uparrow,-\frac{1}{2}} + \ket{\downarrow,-\frac{1}{2}} + \ket{\downarrow,-\frac{3}{2}} \Big) \,,
\label{starg32mt}
\eeq
which is constructed to be non-reducible to a spin \((\frac{1}{2} - \frac{1}{2})\) state via local unitary operations on the qudit (see Supplementary Material). This guarantees that the we exploit the full qubit--qudit structure.
For this state, we numerically determine the observables \( A, A', B, B' \) that maximize the violation of the Bell inequality, yielding a theoretical upper bound:
\begin{equation}
 \langle \psi | \hat{O}_{\rm Bell} | \psi \rangle = 2.64575 \,.
\end{equation}
This benchmark sets the maximum violation achievable in our scenario and is used to evaluate the results of the implemented protocol.\\
{\it Molecular spin dimer to test CHSH inequality --} The entangled state $\ket{\psi}$ in Eq. \eqref{starg32mt} can be experimentally realized using a Yb(trensal) molecule \cite{hansen}, which hosts an electronic effective spin 1/2 coupled by hyperfine interaction to a nuclear spin 5/2. We exploit the electronic spin to encode the qubit and four nuclear spin states to encode the  spin $\frac{3}{2}$-qudit. 
The molecule is shown in Fig. \ref{figura_completa}-(a) and described by the following Hamiltonian:
\begin{align}
H &= g_S \mu_B B_z S_{z} + g_I \mu_N B_z I_{z} + A_{\perp} (S_{x} I_{x} + S_{y} I_{y}) \, + \notag \\
  &\quad + A_{\parallel} S_{z} I_{z} + p I_{z}^2 \, ,
\label{ham31}
\end{align}
where $g_S = (2.9, 2.9, 4.3)$ represents an anisotropic electronic $g$-factor, $g_I = -0.2592$ is the nuclear $g$-factor, $A_{\parallel} = -883$ MHz and $A_{\perp} = -628$ MHz are the hyperfine coupling constants along the $z$-axis and in the perpendicular plane, respectively, and $p = -66$ MHz  represents the axial nuclear quadrupole interaction. 
A level diagram as a function of the external field is reported in  \ref{figura_completa}-(b). 
For the simulation, we consider an external static magnetic field $B_z=0.3$ T applied along the $z$-axis.
At this field, the eigenstates are practically factorized into electronic and nuclear spin  components and hence represent the good starting point to test CHSH inequalities.  
Nonetheless, a tiny electro-nuclear mixing of the wavefunction yields a sizable enhancement of the matrix elements of nuclear spin transitions, enabling fast manipulations of the nuclear states. 
At this optimized magnetic field, we find a remarkably long nuclear spin coherence times of $T_{2n} \gtrsim 500$ $\mu s$.
These values have been experimentally measured with pulsed EPR and broadband NMR at $B_z = 0.3$ T in single crystal samples of $^{173}$Yb(trensal) with a 0.05$\%$ dilution - the best compromise for having a strong enough echo signal and negligible dipolar inter-molecular interactions. The results are shown in the Supplementary Information along with additional experimental  details. The nuclear coherence time $T_{2n}$ was measured for four transitions among the hyperfine levels of Yb(trensal) and fitted to an exponential function, consistently with the Lindblad master equation used in our numerical simulations (vide infra). We considered the shortest $T_{2n}=560 \; \mu$s obtained from these fits, in order to include the effect of pure dephasing in the less favorable conditions.
 \\
The electronic spin coherence time in the same conditions of dilution and external field was measured by pulse EPR, resulting in $T_{2e} \sim 2.4$ $\mu s$ at 2.7 K (see Supplementary Information).

{\it Pulse sequence and demonstration of violation of the CHSH inequality-} The target state $\ket{\psi}$ in Eq. \eqref{starg32mt} is obtained starting from the ground state  $\ket{\downarrow, -\frac{5}{2}}$ of the Hamiltonian in Eq. \eqref{ham31} and applying 
suitable sequences of electromagnetic pulses  to implement single-qubit, single-qudit, and joint qubit-qudit gates \cite{review} (see Supplementary Information for details). These are realized via resonant square pulses, whose frequency and duration define which transition is addressed and the rotation angle. 
Numerical simulations of the whole pulse sequence, including the effect of pure dephasing, are performed by solving the Lindblad master equation:
\begin{equation}
\frac{d{\rho}}{dt} = -i[{H}, {\rho}] + \sum_{j} \gamma_j \left( L_j \rho L_j^\dagger - \frac{1}{2} \left\{ L_j^\dagger L_j, \rho \right\} \right),
\label{eq:lindblad}
\end{equation}
where $\gamma_j$ denotes the dephasing rate acting on site $j$ and $L_j$ are the corresponding jump operators \cite{lindblad1976}.
At low temperatures, these systems exhibit very long relaxation times and pure dephasing is by far the dominant noise effect \cite{blueprint,noise}. Hence, the relevant jump operators in Eq. \eqref{eq:lindblad} are to $S_z$ and $I_z$ with corresponding rates $\gamma_j = 1/T_{2e}$ and $1/T_{2n}$. \\
At the end of the simulation, we compute the fidelity $\mathcal{F} = \bra{\psi} \rho \ket{\psi}$ between the final  density matrix $\rho$ and the target state. 
The fidelity is optimized by tuning the pulse amplitude $B_1$, 
balancing spectral selectivity and decoherence.
Indeed, the pulse duration is proportional to $1/B_1$ and hence increasing $B_1$ on the one hand reduces the effect of decoherence, on the other hand can yield leakage towards transitions occurring at similar frequencies. 
In particular, we obtain $\ket{\psi}$ through five pulses: one microwave addressing a qubit transition for a specific nuclear spin state and 4 radiofrequency pulses addressing nuclear spin transitions. \\
To optimize the pulse sequence and test its robustness to noise, we have studied the fidelity in the construction of $\ket{\psi}$ as a function of $B_1$ and of the coherence times. First, we note that experimentally measured values of $T_{2n}$ are much longer than the pulse sequence duration, typically lasting 0.5-1 $\mu$s for the optimal $B_1$. Hence, we fixed $T_{2n}$ at the experimentally measured value of $560 \; \mu$s and compute  $\mathcal{F}$ as a function of $T_{2e}$ and $B_1$. By proper choice of $B_1$, fidelities above $0.95$ were obtained for all values of $T_{2e}$. Results are shown in Fig.~\ref{figura_completa}(c); in particular, in the construction of the entangled state $\ket{\psi}$, for all decoherence scenarios a maximum appears and shifts to higher $B_1$ as $T_{2e}$ decreases.    
In particular, a fidelity above 0.94 is obtained using the measured $T_{2e}=2.4 \; \mu$s [highlighted by a thick box in Fig. \ref{figura_completa}-(c)].

The expectation value of the Bell operator in Eq.~\eqref{bellexp} can be equivalently reformulated in terms of a more experimentally accessible set of four expectation values of diagonal spin operators, $S_z$ for the qubit and  $D_j$ for the qudit, after rotating the state via local unitaries \footnote{Note that any evolution due to the hyperfine coupling term in Eq. \eqref{ham31} is ineffective on the measured diagonal observables.}.
It can thus be expressed as
\begin{equation}
\langle \psi | \hat{O}^{(\text{max})}_{\rm Bell} | \psi \rangle = O_{11} + O_{12} + O_{21} - O_{22} \,,
\end{equation}
where each $O_{ij}$ corresponds to a joint measurement on the rotated state $\rho_{A_i,B_j} = (U_{A_i} \otimes U_{B_j}) \ket{\psi}\bra{\psi} (U_{A_i}^\dagger \otimes U_{B_j}^\dagger)$, with $i,j = 1,2$. The four expectation values are 
\begin{equation}
O_{ij} = \mathrm{Tr} \left[ (-2S_z \otimes D_j)\, \rho_{A_i,B_j} \right],
\end{equation}
with
\begin{equation}
D_1 = \mathrm{diag}(-1,1,1,1), \qquad D_2 = \mathrm{diag}(1,-1,1,1) \,.
\label{defDmt}
\end{equation}
Direct implementation of the unitary operations $(U_{Ai} \otimes U_{Bj})$ via hand-optimized pulses yielded insufficient fidelities due to large leakage. To overcome this, we employed GRAPE optimization~\cite{nori2012,nori2013,lambert2024qutip5quantumtoolbox}, which designs optimal control pulses by discretizing control fields into \( N \) time segments, each with constant amplitude. The procedure dynamically optimizes the control fields via gradient ascent maximization of the process fidelity \(\mathcal{F}_U = |\langle U_{\text{target}}|U(T)\rangle|^2\) between the target unitary
$U_{target}=U_{A}\otimes U_{B}$ and the computed time-evolution operator \( U(T) \) (see Supporting Information). In this way, high-fidelity quantum operations can be achieved, balancing computational efficiency and experimental robustness \cite{PhysRevA.69.022319, de_Fouquieres_2011, glaser_2015}.
The control fields were constrained to experimentally realistic values of the amplitudes (maximum $B_1=7.5$ mT) and the frequencies (not exceeding 800 MHz), consistent with typical NMR conditions.  
Within these realistic boundaries, the protocol successfully prepares each of the four rotated states \(\rho_{A_i,B_j}\) in a time of approximately \(1\,\mu\text{s}\).
Remarkably, we obtain values of $\langle \hat{O}^{(\mathrm{max})}_{\mathrm{Bell}} \rangle$ that consistently violate the Bell inequality across all tested decoherence regimes (dashed red line in Fig.~\ref{figura_completa}-(d)) and in particular for $T_{2e}=2.4 \; \mu$s (red circle).
These results demonstrate the robustness of the observed quantum correlations in our qubit–qudit platform in a realistic simulation of the experimental protocol. Remarkably, the scheme for demonstrating violation of CHSH inequalities does not require projective measurements but only ensemble expectation values. 
All in all, these findings pave the way to an actual implementation of the procedure with a broadband nuclear magnetic resonance setup on available molecular crystals of Yb(trensal).  \\

\begin{figure}[!htbp]
\includegraphics[width=0.4\textwidth]{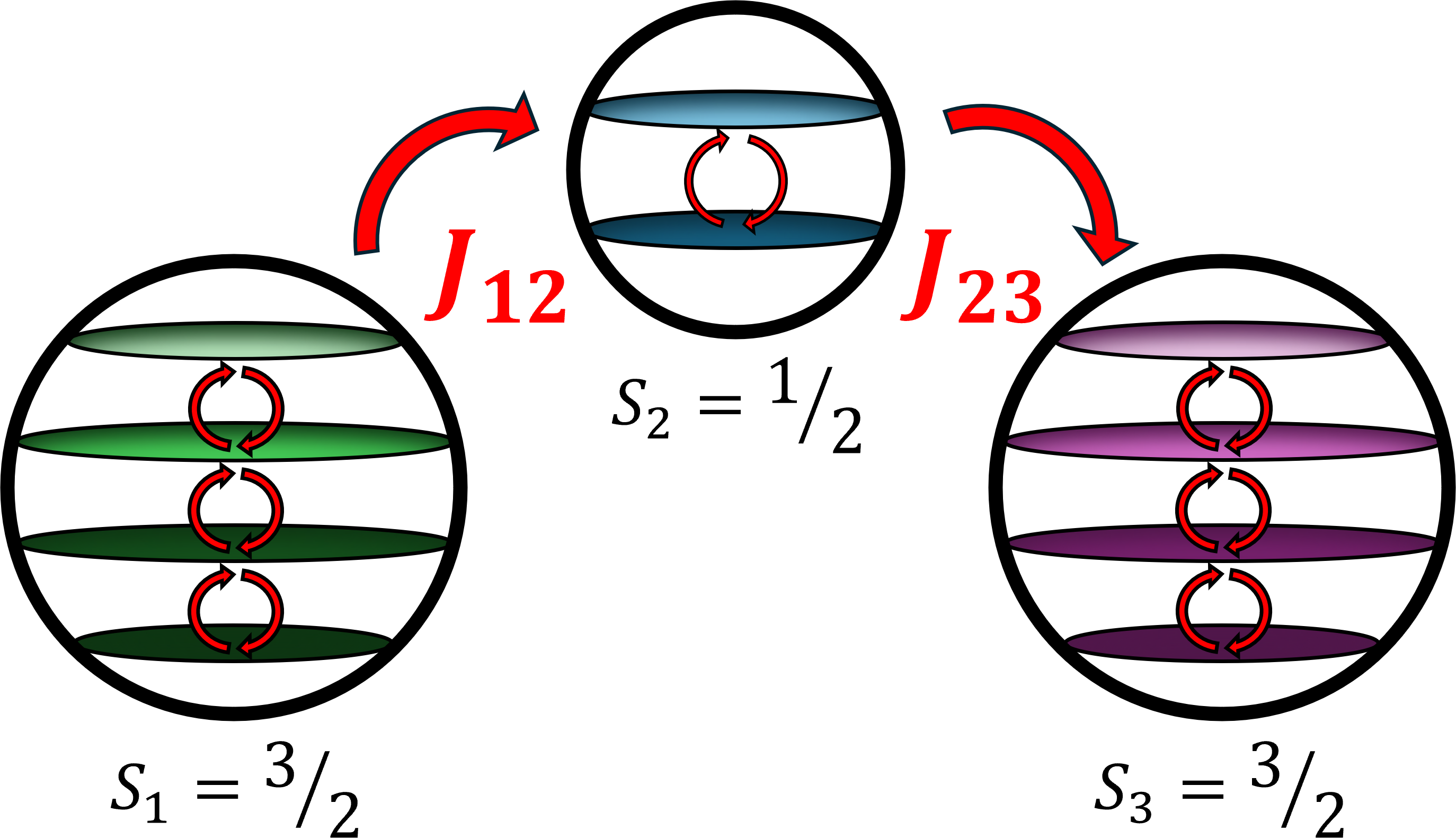} 
\includegraphics[width=0.5\textwidth]{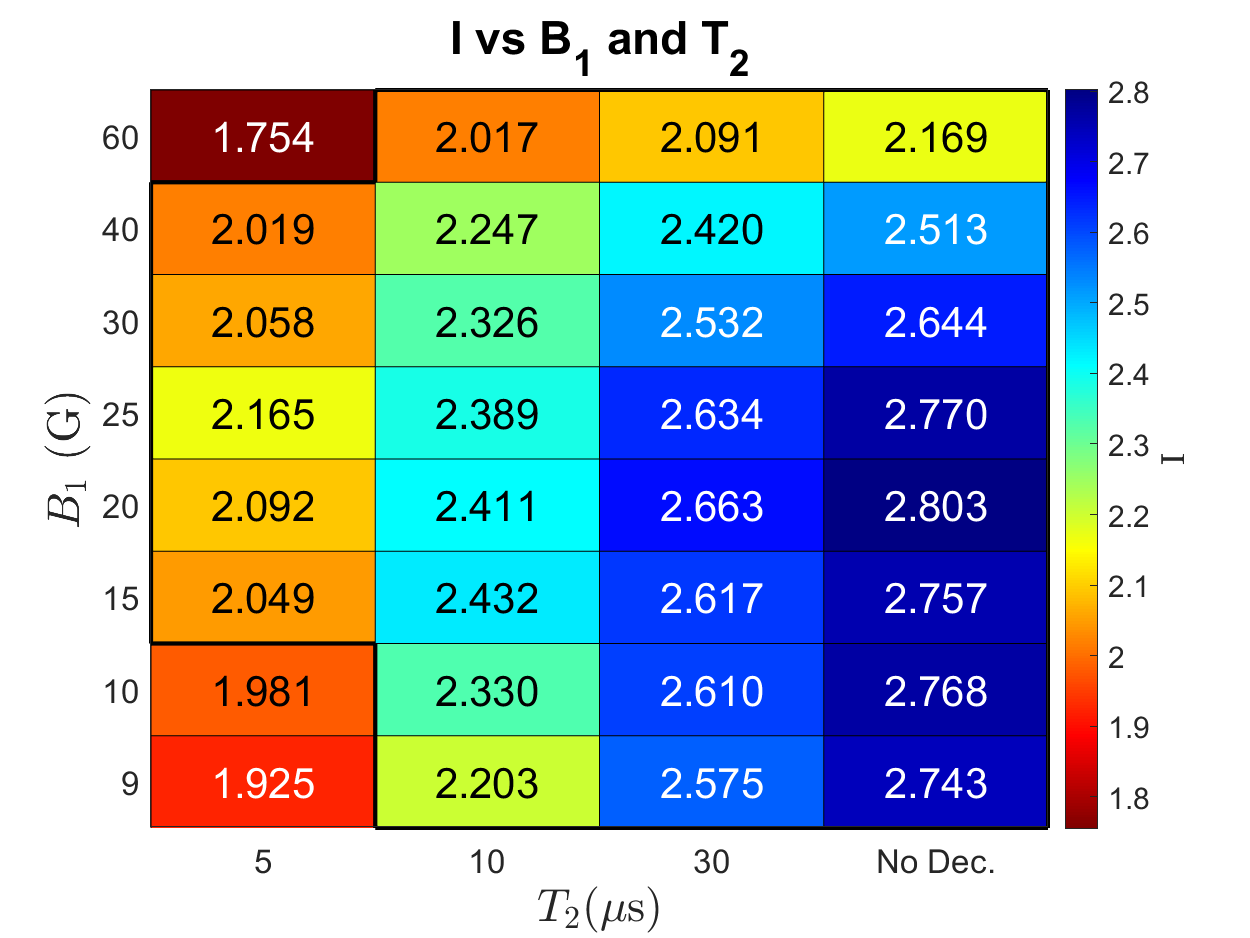}
\caption{Upper panel: schematic representation of the molecular spin trimer proposed to test CGLMP inequality on a pair of four-level qudits. It consists of two spin-$\frac{3}{2}$ qudits ($S_1,S_3$) coupled via isotropic exchange interactions $J_{12}, J_{23}$ to a two-levels switch ($S_2$). A static magnetic field $B_0 = 1.3$ T is applied along the $z$ direction.
Lower panel: value of the Bell functional $I$ as a function of the driving field amplitude $B_1$ and of the dephasing time $T_2$ of each qudit. Violation of the CGLMP inequality ($I>2$) occurs in a wide range of realistic  parameters (thick line). }
  \label{figura1}
\end{figure}

{\it Extension to a two-qudits dimer-} 
In the previous proof-of-concept experiment the coupling between the two parties was always-on, although ineffective in the proposed working conditions. 
We now propose an alternative platform in which the entanglement between the qudits is turned on and off when required, through a switchable coupling mechanism mediated by a spin-$\frac{1}{2}$ ancilla \cite{chiesa2024}. In principle, this system can allow space-like separation. Moreover, we now explore a fully symmetric two-qudit configuration, focusing on an entangled state of two spin-$\frac{3}{2}$ qudits. This mediated interaction allows us to turn on the coupling only when required, enabling switchable  entangling operations. 
A possible realization of this \big(\(\frac{3}{2} - \frac{1}{2} - \frac{3}{2}\)\big) molecular spin trimer is represented by two Cr$^{3+}$ ions in a distorted octahedral environment ($S=3/2$) \cite{AbragamBleaney}, linked through an interleaved  Yb (trensal)  \cite{quditqubit, Hussain}. The latter behaves at low energy as an effective spin doublet. 
The spin Hamiltonian of the system (sketched in Fig.~\ref{figura1}) is:
\begin{align}
&\mathcal{H}=\mu_B B_z \big( g_1^z S_{1}^{z}+g_2^z s_{2}^{z}+g_3^z S_{3}^{z} \big) \, +  \nonumber \\
&\, +J_{12} \big(S_{1}^{x}s_{2}^{x}+S_{1}^{y}s_{2}^{y}+S_{1}^{z}s_{2}^{z} \, \big) \, + \nonumber \\
&+J_{23} \big (s_{2}^{x}S_{3}^{x}+s_{2}^{y}S_{3}^{y}+s_{2}^{z}S_{3}^{z} \big) \, + \nonumber \\
&+  D_1 S_{1}^{z \, 2}+D_3 S_{3}^{z \, 2} , 
\label{hamilt}
\end{align}
where $s_{{2}}^{\alpha}$, $\alpha = (x,y,z)$, is a spin $1/2$ operator associated ancilla  and  $S_{i}^{\alpha}$, $i = 1,3$ is a spin $3/2$ operator. 
An external magnetic field $B_z=1.3$ T is applied along the $z$-axis, activating the Zeeman terms, which are proportional to  $g_1=2.0$,  ${\bf g}_2=(2.9,2.9,4.3)$  and $g_3=1.95$ respectively. 
Additionally, the system includes two isotropic exchange terms, $J_{12}=5\cdot10^{-3}$ meV and $J_{23}=3\cdot10^{-3}$ meV between neighboring spins, as well as axial zero-field splitting (ZFS) parameters $D_1=-3\cdot10^{-2}$ meV and $D_3=-2\cdot10^{-2}$ meV. All these parameters are reasonable for Cr and Yb subunits and the weak exchange can be achieved in Cr-Yb molecules. \footnote{The weak exchange interaction values between Cr and Yb used in this study are consistent with those reported in the literature for analogous Cr-Lanthanide complexes, where magnetic interactions are generally considered very weak. For example, for the Cr-Gd coupling, a $J = +0.02~\text{cm}^{-1}$ (weakly ferromagnetic) was determined in \cite{sanada}}
Such a configuration enables a direct application of the Collins-Gisin-Linden-Massar-Popescu (CGLMP) inequality~\cite{Collins_2002}, which is designed to test the non-classicality of correlations in high-dimensional systems. The inequality is constructed from a functional of the joint outcome probabilities:
\begin{align}
\label{eq666}
& I=[P(A_1=B_1)+P(B_1=A_2+1)+P(A_2=B_2) \, + \\ \nonumber
&+P(A_1=B_2)] 
-[P(A_1=B_1-1)+P(A_1=B_2)\, + \\ \nonumber
&+P(A_2=B_2-1)+P(B_2=A_1-1)] \, + \\ \nonumber
&+ \frac{1}{3}[P(A_1=B_1+1)+P(B_1=A_2+2)+P(A_2=B_2+1) \, + \\ \nonumber
&+P(B_2=A_1+1)] 
-\frac{1}{3}[P(A_1=B_1-2) \, +\\ \nonumber
&+P(B_1=A_2-1)+P(A_2=B_2-2)+P(B_2=A_1-2)], 
\end{align}
where \( P(X_{\alpha}=Y_{\beta}) \) indicates the probability that the outcomes of the local observables \( X_\alpha \) and \( Y_\beta \) coincide. In our notation, \( A_1 \) and \( A_2 \) denote the outcomes of two measurements performed on the first subsystem, and \( B_1 \), \( B_2 \) those on the second subsystem. The evaluation of the functional \( I \) is analogous to that of the Bell operator \( \hat{O}^{(\mathrm{max})}_{\mathrm{Bell}} \) discussed above, and violations of local hidden-variable models are observed for \( I > 2 \). \\
To test the CGLMP inequality with maximal violation, we target the maximally entangled state
\begin{equation}
    \ket{\psi} = \frac{1}{\sqrt{4}} \sum_{m=0}^{3} |m\rangle_A \otimes |m\rangle_B \, ,
\end{equation}
and prior to measurements we apply four unitary transformations \( U^{(A)}_i \) and \( U^{(B)}_j \) ($i,j = 1,2$)  \cite{Polozova,Collins_2002} with matrix elements 
\begin{equation}
    [U^{(A)}_i]_{kl} = \frac{1}{\sqrt{4}} \, e^{\frac{2\pi i\, l}{4}(\alpha_i + k)}, \qquad
    [U^{(B)}_j]_{kl} = \frac{1}{\sqrt{4}} \, e^{\frac{2\pi i\, l}{4}(\beta_j - k)}.
    \label{unitaries}
\end{equation}
Here \( \alpha_1 = 0 \), \( \alpha_2 = \frac{1}{2} \), \( \beta_1 = \frac{1}{4} \), and \( \beta_2 = -\frac{1}{4} \) \cite{Polozova}. \\
Finally, expectation values of $S_z$ are computed. 
In the same spirit of the previous section, we test the protocol on the proposed molecular setup by performing numerical simulations of the whole pulse sequence in presence of decoherence. To this goal, we initialize the system in the ground state
$\ket{-3/2}_1 \otimes \ket{-1/2}_2 \otimes \ket{-3/2}_3$ and we drive it, through a sequence of single- and two-qudit gates (see Supplementary Material), to the target state
\begin{align}
\label{eq42}
\ket{\psi} 
&= \tfrac{1}{2} \big(
\ket{-\tfrac{3}{2}, -\tfrac{3}{2}} +
\ket{-\tfrac{1}{2}, -\tfrac{1}{2}} +
\ket{+\tfrac{1}{2}, +\tfrac{1}{2}} + \notag \\
&\ket{+\tfrac{3}{2}, +\tfrac{3}{2}} \big)_{13} \otimes \ket{-\tfrac{1}{2}}_2 \, .
\end{align}
corresponding to a maximally entangled state of the two qudits. 
The effective interaction between the two qudits is  controlled via a central ancilla qubit that acts as a switchable mediator. When the ancilla is frozen in its ground state, no entangling interaction takes place. Conversely, by a conditional excitation of the ancilla depending on the state of both qudits we implement a two-qudit controlled-phase gate, which turns on qudit-qudit entanglement \cite{chiesa2024, Switch, FERRANDOSORIA2016727, santini2011}. After enabling the entangling interaction, the ancilla is brought back to its ground state, resulting in a final state where it is factorized from the two entangled qudits.

The sequence is simulated by solving the Lindblad master equation, exploring realistic pulse amplitudes and dephasing times to identify the optimal trade-off between leakage and decoherence. 
Remarkably, the much shorter coherence time of the Yb ancilla ($T_2=1\;\mu$s) has only a limited role, because the ancilla is brought to a superposition only during the implementation of two-qudit gates.
Optimized fidelities with the exact state are above 0.91 for each of the considered $T_2$ values of the qudits (between 5 and 30~$\mu$s). In all considered dephasing scenarios, the Bell inequality threshold of \(I = 2\) is clearly exceeded, reaching values up to \(I = 2.61\) for \(T_2 = 30~\mu\mathrm{s}\), as shown in Fig.~\ref{figura1}.\\

In summary, this work demonstrates the significant potential of molecular nanomagnets as a robust and versatile platform for exploring the quantum properties of matter, particularly within the less-explored territory of multi-dimensional quantum systems ({\it qudits}). 
In particular, we have shown by numerical simulations based on real parameters and on a feasible experimental setup that Yb(trensal) electro-nuclear spin dimer can be used to probe entanglement by testing generalized Bell inequalities. Moreover, we have proposed an extension of the scheme to a spin-qudit dimer in which the entanglement between the two qudits can be dynamically switched on and off by exciting an interposed qubit switch. These results place molecular nanomagnets as strong candidates not only for applicative issues but also for understanding fundamental quantum phenomena such as qudit-qudit entanglement. An interesting evolution of the present study involves the derivation and test of inequalities for the qubit-qudit case beyond the CHSH ones, in particular based on non-dichotomic measurements.\\

{\bf Acknowledgments --} The Authors are pleased to thank J. A. Casas and P. Santini for fruitful discussions. \\
The Authors acknowledge financial support by a project funded under the National Recovery and Resilience Plan (NRRP), Mission 4 Component 2 Investment 1.3 - Call for tender No. 341 of 15/03/2022 of Italian Ministry of University and Research funded by the European Union – NextGenerationEU, award number PE0000023, Concession Decree No. 1564 of 11/10/2022 adopted by the Italian Ministry of University and Research, CUP D93C22000940001, Project title "National Quantum Science and Technology Institute" (NQSTI), spoke 2.

\newpage

\onecolumngrid
\appendix

\section{General unitary inequivalence issue between  $\big(\frac{1}{2}-S\big)$ and $\big(\frac{1}{2}-\frac{1}{2}\big)$ states}
 \label{gis}
First, we deal with the general issue to determine when a (normalized) state 
\beq
\ket{\psi^{\prime}} =\sum_{s = -S}^{S} \alpha_{s} \ket{\frac{1}{2}, s} + \sum_{s^{\prime} = - S}^{S} \beta_{s^{\prime}} \ket{-\frac{1}{2}, s^{\prime}} \,
\label{vecgen}
\eeq
 is equivalent (or not)  to at least a $\big(\frac{1}{2}-\frac{1}{2}\big)$ state  via a local unitary transformation 
 \beq
\tilde{U} = {\bf I}_{2 \mathrm{x} 2} \times U \, ,
\label{utransf}
\eeq
where $U \in U(4) = \frac{SU(4) \times U(1)}{Z_4}$  \Big($\frac{\{\}}{\{\}}$ denoting a quotient space). In the latter formula, $U(4)$ are the $4 \, \mathrm{x} \, 4$ unitary matrix transformations,  $SU(4)$ its subset with determinant equal to 1, and  $Z_4 =\{1, e^{i \frac{\pi}{4}} , e^{i \frac{\pi}{2}} , e^{i \frac{3 \, \pi}{4}} \} \, {\bf 1}$ indicates the center of $SU(4)$ \cite{georgi}. This transformation, being
 local, does not change the entanglement between the two
 parties.\\

 We consider first the case $S = \frac{3}{2}$.
 For our purposes, it is convenient to begin from the state\\
  (again staying, without any loss of generality, in the $\{\frac{3}{2} , -\frac{3}{2} \}$ subspace of the $S= \frac{3}{2}$ manifold; it can be checked that all the other subset choices are unitary equivalent, via permutations, to the present one)
 \beq
 \ket{\psi} = a \, \ket{\frac{1}{2}, \frac{3}{2}} + b \, \ket{-\frac{1}{2}, - \frac{3}{2}} \, , \quad |a|^2 + |b|^2 = 1 \, ,
 \label{psifin}
 \eeq
 that can be conveniently rewritten, in the $H_{\frac{1}{2}} \otimes H_{\frac{3}{2}}$ space, as
\beq
\ket{\psi} =  \big(a,0,0,0,0,0,0, b \big)^T \, .
\label{psi3}
\eeq 
Equivalently to the requirement on the state in Eq. \eqref{vecgen}, we want to discuss if and when from $\ket{\psi}$ in Eq. \eqref{psi3}, a generic, still normalized, state
\beq
\ket{\psi^{\prime}} =  \big(\alpha_1,\alpha_2,\alpha_3,\alpha_4,\beta_5,\beta_6,\beta_7, \beta_8  \big)^T 
\label{psi4}
\eeq
is obtained by a local unitary transformation as in Eq. \eqref{utransf}.\\

For our task, it is convenient to consider first that, referred to Eq. \eqref{utransf},
\begin{align}
\centering
& U \big(a,0,0,0)^T = a \, \big(U_{11}, U_{12}, U_{13},U_{14})^T= \big(\alpha_1,\alpha_2,\alpha_3,\alpha_4\big)^T \, , \nonumber \\
& U \big(0,0,0,b)^T =b\,\big(U_{41}, U_{42}, U_{43},U_{44}\big)^T = \big(\beta_5,\beta_6,\beta_7,\beta_8\big)^T \, .
\label{condU}
\end{align}
From this equation and due to the unitarity of $U$, it emerges immediately that a first requirement is
\beq
 \big(U_{11}, U_{12}, U_{13},U_{14}\big) \cdot \big(U_{41}, U_{42}, U_{43},U_{44}\big)^T = \big(\alpha_1,\alpha_2,\alpha_3,\alpha_4\big) \cdot \big(\beta_5,\beta_6,\beta_7,\beta_8\big)^T=
 \big(a,0,0,0 \big) \cdot \big(0,0,0, b \big)^T= 0 \, .
 \label{condort}
\eeq
Therefore, all the states as in Eq. \eqref{vecgen} \emph{without} this property are \emph{not reducible} in the form as in Eq. \eqref{psifin} by a local unitary transformation as in Eq. \eqref{utransf}.  \\

 If the condition  in Eq. \eqref{condort} is fulfilled, a second pair of necessary conditions (still related to the unitarity of $U$) is that
\beq
\mathrm{norm} \big[ \big(a,0,0,0)^T \big] = \mathrm{norm} \big[ \big(\alpha_1,\alpha_2,\alpha_3,\alpha_4)^T \big]  = a \, ,
\label{cond1}
\eeq
and
\beq
\mathrm{norm} \big[ \big(0,0,0,b)^T \big] =  \mathrm{norm} \big[ \big(\beta_5,\beta_6,\beta_7,\beta_8)^T \big] = b \, .
\label{cond2}
\eeq
These conditions can be fulfilled choosing properly (and up to two phases) $a$ and $b$ in Eqs. \eqref{psifin} and \eqref{psi3}. 
In this way, exploiting again Eq. \eqref{condU}, the elements of $U$ reported in the same equation can be fixed. We can choose, for instance
\beq
U_{11} = \frac{\alpha_1}{a} \, ,
\label{fix}
\eeq
and similarly for the other $U_{ij}$ unitary matrix elements  in Eq. \eqref{condU}. Notice that, due to the conditions in Eqs. \eqref{cond1} and \eqref{cond2}, $|U_{ij}|<1$ for the same elements fixed as in Eq. \eqref{fix}.

After this first choice, only $4^2- 2 \cdot 4 = 8$ complex coefficients  $\{U_{21}, U_{22}, U_{23},U_{24}, U_{31}, U_{32}, U_{33},U_{34}\}$,
corresponding to $2 \, \big(4^2- 2 \cdot 4\big) = 16$ real coefficients, still remain to be fixed for $U$. In principle, this can be done by the $4^2 = 16$ conditions for $U$ to be unitary
\beq
\sum_k U^*_{ik} U_{kj} = \delta_{ij} \, .
\label{condU2}
\eeq
However, 3 conditions  in this set are already fulfilled \big(from setting $\{i,j\}=\{1,4\}$ in Eq. \eqref{condU2} and also encoded in Eqs. \eqref{condort},  \eqref{cond1}, and \eqref{cond2}\big); therefore $4^2-3=13$ conditions remain.
The central point is now that these conditions are not sufficient to fix entirely the 16 real coefficients mentioned above.

 Therefore, if (and only if !) Eq. \eqref{condort} is fulfilled,  the $U$ \big($\tilde{U}$, cfr. Eq. \eqref{utransf}\big) in question can be found, and it is not even unique.
In turn, this means that, iff Eq. \eqref{condort} is fulfilled by a spin $\big(\frac{1}{2}-\frac{3}{2} \big)$ state, the same state can be  reduced by a local unitary transformation as in Eq.  \eqref{utransf} to a $\big(\frac{1}{2}-\frac{1}{2} \big)$ state as in Eq. \eqref{psifin}: all the possibilities are covered by varying $a$ and $b$ in the same equation. \\
Indeed, it is known that a unitary $n \times n$ matrix, with $n = 2 S + 1$, is defined by $n^2$ conditions, as in Eq.~\eqref{condU2}. Again, two conditions, corresponding to $\{i,j\} = \{1,n\}$ in the set of Eq.~\eqref{condU2}, are already fulfilled due to the equivalent constraints given in Eqs.~\eqref{condort}, \eqref{cond1}, and \eqref{cond2}. Therefore, $n^2 - 3$ conditions can be used to partially fix the $2(n^2 - 2n) \geq n^2 - 3$ real coefficients in $U$ that remain after imposing the equivalent conditions to those in Eq.~\eqref{condU}. The equality is saturated when $n = 3$, that is, for $S = 1$.
This number of constraints is always greater than or equal to $n^2 - 3$; the equality is saturated only for $n = 3$, i.e., for $S = 1$.

Therefore, again the searched $U$ in Eq. \eqref{utransf} always exists (not even unique) iff Eq. \eqref{condort} is fulfilled.

\section{Maximization of $\langle \psi | \hat{O}_{\rm Bell} | \psi \rangle$ for the $\big(\frac{1}{2}-\frac{3}{2}\big)$ case}
\label{appmax}
The next step is to maximize, following \cite{moreno2024}, the expectation value of an operator 
with the form
\begin{equation}
 \hat{O}_{\rm Bell} = A \otimes (B+B') + A'\otimes (B-B') \ 
 \label{defbell}
\end{equation}
on the  pure $\big(\frac{3}{2}-\frac{1}{2}\big)$ state
\beq
\ket{\psi} = \frac{1}{2} \, \Bigg(\ket{\frac{1}{2},\frac{3}{2}} + \ket{\frac{1}{2},-\frac{1}{2}} + \ket{-\frac{1}{2},-\frac{1}{2}} + \ket{-\frac{1}{2},-\frac{3}{2}} \Bigg) \, .
\label{starg32supp}
\eeq
For this task, we follow closely the formalism outlined in \cite{moreno2024}.
We first express the density matrix $\rho =  \ket{\psi} \bra{\psi}$ as follows:
\beq
\rho = \frac{1}{2} \, \Bigg[{\bf 1}_{2\times 2} \times \tilde{\sigma}_0 + \sum_{i = 1}^3 \sigma_i \times \tilde{\sigma}_i \Bigg] \, ,
\label{rhopar}
\eeq
where $\sigma_i$ are the Pauli matrices,
\beq
\tilde{\sigma}_{\alpha} = \mathrm{Tr}_{| H_{2}} \Big[\big(\sigma_{\alpha} \otimes {\bf I}_{4 \mathrm{x} 4} \big) \, \rho \Big] \, ,  \, \quad \alpha = 0, \dots , 4 \, ,
\eeq
$\mathrm{Tr}_{| H_{2}}$ being the trace over the the spin-$\frac{1}{2}$ subsystem,  and $\sigma_0 = {\bf I}_{2 \mathrm{x} 2} $.
Explicitly, for $\ket{\psi}$ in Eq. \eqref{starg32supp}:
\beq
\tilde{\sigma}_0 =  \frac{1}{4} \, \left(
\begin{array}{cccc}
 1 & 0 & 1 & 0 \\
 0 & 0 & 0 & 0 \\
 1 & 0 & 2 & 1 \\
 0 & 0 & 1 & 1 \\
\end{array}
\right) \, , \, \,
\tilde{\sigma}_1 =  \frac{1}{4} \, \left(
\begin{array}{cccc}
 0 & 0 & 1 & 1 \\
 0 & 0 & 0 & 0 \\
 1 & 0 & 2 & 1\\
 1 & 0 & 1 & 0 \\
\end{array}
\right) \, , \, \,
 \tilde{\sigma}_2 =  \frac{i}{4} \, 
 \left(
\begin{array}{cccc}
 0 & 0 & 1 &1 \\
 0 & 0 & 0 & 0 \\
 -1 & 0 & 0 & 1 \\
 -1 & 0 & -1 & 0 \\
\end{array}
\right)
  \, , \,  \,
\tilde{\sigma}_3 =  \frac{1}{4} \,
\left(
\begin{array}{cccc}
 1 & 0 & 1 & 0 \\
 0 & 0 & 0 & 0 \\
 1 & 0 & 0 & -1 \\
 0 & 0 & -1 & -1 \\
\end{array}
\right) \, .  
\eeq
At this point, the maximum value of $\langle \psi | \hat{O}_{\rm Bell} | \psi \rangle$, on the set of all the possible Bell operators $ \hat{O}_{\rm Bell} $ is given (for any $S$) by
\begin{equation}
\langle \psi | \hat{O}_{\rm Bell} | \psi \rangle_{\rm max}
= 
2\, \max_{{\cal R}} 
\left[
            \left(\sum_{i=1}^{2S+1= 4} |\lambda_i^{(1)}({\cal R})| \right) ^2  +
           \left( \sum_{i=1}^{2S+1=4} |\lambda_i^{(2)}({\cal R})| \right) ^2
\right]^{1/2} \, ,
\label{max}
\end{equation}
where $\lambda_i^{(1,2)} (\mathcal{R})$ are the eigenvalues of the rotated matrices 
$(\mathcal{R} \vec{\tilde{\sigma}})_1$ and $(\mathcal{R} \vec{\tilde{\sigma}})_2$, with $\mathcal{R}(\alpha, \beta, \gamma)$ denoting the $3 \times 3$ real matrix representation of a rotation acting on spin-1 vectors.
Therefore, the maximum of Eq. \eqref{max} must be evaluated on the set of values for the three angles $(\alpha, \beta , \gamma)$.

The $A, A', B, B'$ observables that realize the maximum CHSH-violation are obtained as follows. In general \cite{moreno2024}, once we have determined the matrices $({\cal R}\vec{\tilde{\sigma}})_1$, $({\cal R}\vec{\tilde{\sigma}})_2$ that maximize Eq. (\ref{max}), one  sets
\beq
B=U_1^\dagger D_1 U_1 \, , \quad
B^{\prime}=U_2^\dagger D_2 U_2 \, ,
\label{paramB}
 \eeq
where $U_{1,2}$ are the diagonalizing unitary matrices of $({\cal R}\vec{\tilde{\sigma}})_i$, i.e. $U_i ({\cal R}\vec{\tilde{\sigma}})_i U_i^\dagger={\rm diag}\,(\lambda_a^{(i)})$, and $D_i={\rm diag}({\rm sign}\,[\lambda_a^{(i)}])$ (here $i = 1,2$). This means that $B$ and $B^{\prime}$ get simultaneously diagonalized with $({\cal R}\vec{\tilde{\sigma}})_i$.\\

In our case, it turns out that various maximizing solutions exist, yielding the same maximal value. 
One maximal solution of the described numerical procedure is 
$(\alpha, \beta, \gamma) = (0,\pi,0)$,
\beq
B =\frac{1}{3} \,\left(
\begin{array}{cccc}
2 & 0 & -2 & -1 \\
 0 & 3 & 0 & 0 \\
 -2 & 0 & -1 & -2 \\
 -1 & 0 & -2 & 2 \\
\end{array}
\right)
\, , \quad 
B^{\prime} =  \left(
\begin{array}{cccc}
\frac{1}{3} & 0 & -\frac{1}{3}+\frac{\sqrt{3}}{3} i & \frac{1}{3}\, +\frac{\sqrt{3}}{3} i \\
 0 & 1. & 0 & 0 \\
 -\frac{1}{3}-\frac{\sqrt{3}}{3} i & 0 & \frac{1}{3} & -\frac{1}{3}+\frac{\sqrt{3}}{3}i \\
\frac{1}{3}\, -\frac{\sqrt{3}}{3} i & 0 & -\frac{1}{3}-\frac{\sqrt{3}}{3} i & \frac{1}{3} \\
\end{array}
\right)
 \, ,
\eeq
 with
\beq
\langle \psi | \hat{O}_{\rm Bell} | \psi \rangle = 2.64575 \, .
\label{fval2.02_app}
\eeq
This value is still significantly larger than the threshold $2$ required 
to witness the violation of the CHSH inequalities for the operators in Eq. \eqref{defbell} \cite{3hor,moreno2024}. 

The operators $B$ and $B^{\prime}$ are parametrized as 
\beq
B=U_{B1}^\dagger D_1 U_{B1} \, , \quad
B^{\prime}=U_{B2}^\dagger D_2 U_{B2} \, ,
\label{paramB}
 \eeq
 with
\beq
D_1 =  \left(
\begin{array}{cccc}
 -1 & 0 & 0 & 0 \\
 0 & 1 & 0 & 0 \\
 0 & 0 & 1 & 0 \\
 0 & 0 & 0 & 1 \\
\end{array}
\right) \, ,  \quad \quad D_2 =  \left(
\begin{array}{cccc}
 1 & 0 & 0 & 0 \\
 0 & -1 & 0 & 0 \\
 0 & 0 & 1 & 0 \\
 0 & 0 & 0 & 1 \\
\end{array}
\right)
\label{defD}
\eeq
(where we have arbitrarily set $\mathrm{sign} (0) = 1$, as suggested in \cite{moreno2024}), and
\beq
U_{B1} =\left(
\begin{array}{cccc}
 -\frac{1}{\sqrt{6}}& 0 & -\frac{2}{\sqrt{6}} & -\frac{1}{\sqrt{6}} \\
 -\frac{1}{\sqrt{2}} & 0 & 0 & \frac{1}{\sqrt{2}} \\
 \frac{\sqrt{3}}{3} & 0 & -\frac{\sqrt{3}}{3} & \frac{\sqrt{3}}{3}\\
 0 & 1. & 0 & 0 \\
\end{array}
\right) \, ,
\label{defB1a}
\eeq
and
\beq
U_{B2} = \left(
\begin{array}{cccc}
\frac{\sqrt{3}}{6}\, +\frac{1}{2} i & 0 & -\frac{\sqrt{3}}{6}+\frac{1}{2}i & -\frac{\sqrt{3}}{3} \\
 -\frac{\sqrt{3}}{6}+\frac{1}{2} i & 0 & \frac{\sqrt{3}}{6}\, +\frac{1}{2} i & \frac{\sqrt{3}}{3} \\
 -\frac{\sqrt{3}}{3} & 0 & \frac{\sqrt{3}}{3} & \frac{\sqrt{3}}{3} \\
 0 & 0.9998\, -0.0189 i & 0 & 0 \\
\end{array}
\right) \, .
\label{defB2a}
\eeq

From Eq. \eqref{defrB}, we further obtain
\beq
\vec{r}_B =(-1,0,0) \, , \, \, \mathrm{and} \, \, \, \, \vec{r}_{B^{\prime}} = (0, \frac{\sqrt{3}}{2}, 0) \, ,
\eeq
while from Eq. \eqref{defrA}
\beq
\vec{r}_A = (-1.51186,1.30931,0)  \, , \, \, \mathrm{and} \, \, \, \, \vec{r}_{A^{\prime}} = - (1.51186,1.30931,0) \, .
\eeq
Following again Ref. \cite{moreno2024}, the corresponding $A,A'$ observables are obtained as follows.
We define for $B$ the vector
\beq
\vec{r}_B  =  \left(\mathrm{Tr} \left[\tilde{\sigma}_1 B \right], \mathrm{Tr}\left[\tilde{\sigma}_2 B\right], \mathrm{Tr} \left[\tilde{\sigma}_3 B\right]
\right ) \, .
\label{defrB}
\eeq
Analogously $B'$, $A,A'$ are identified by the vectors  
\beq
\hat{r}_A = 2 \, \frac{( \vec{r}_B +  \vec{r}_{B^\prime})}{| \vec{r}_B +  \vec{r}_{B^\prime}|} \, ,  \quad \mathrm{and} \quad  \hat{r}_{A'}  = 2 \,  \frac{(\vec{r}_B -  \vec{r}_{B^\prime})}{|\vec{r}_B -  \vec{r}_{B^\prime}|} \, :
\label{defrA}
\eeq
\beq
A = \frac{1}{2} \sum_i \vec{r}_{A i} \, \sigma_i \, ,
\label{defA}
\eeq
and similarly for $A'$. Notice that $\vec{r}_A $ and  $\vec{r}_{A^{\prime}}$ are normalized to 2 \cite{moreno2024}.
From Eq. \eqref{defA}, in our case it results explicitly
\beq
A = \left(
\begin{array}{cc}
 0. & -0.7559-0.6547 i \\
 -0.7559+0.6547 i & 0. \\
\end{array}
\right) \, , \quad 
A^{\prime} =  \left(
\begin{array}{cc}
 0. & -0.7559+0.6547 i \\
 -0.7559-0.6547 i & 0. \\
\end{array}
\right) \, ,
\label{defA}
\eeq
then $A^{\prime} = A^*$. These matrices are diagonalized by the unitary transformations
\beq
U_{A1} = \left(
\begin{array}{cc}
 0.5345\, -0.4629 i & 0.7071 \\
 -0.5345+0.4629 i & 0.7071 \\
\end{array}
\right) \, , \quad \quad U_{A2} = \left(
\begin{array}{cc}
 0.5345\, +0.4629 i & 0.7071 \\
 -0.5345-0.4629 i & 0.7071 \\
\end{array}
\right) \, :
\label{u31}
\eeq
\beq
A_D = U_{A1}^{\dagger} \, A \,  U_{A1} = -\sigma_3 \, , \quad \mathrm{and}  \quad A^{\prime}_D = U_{A2}^{\dagger} \, A^{\prime} \,  U_{A2} = -\sigma_3 \, .
\label{Adiags}
\eeq
The equality $A_D = A^{\prime}_D$ is expected, since $A$ and $A^{\prime} = A^*$ are unitary equivalent each others: $A^{\prime}  = V^{\dagger} \, A \, V$, with $V = \sigma_1 = V^{\dagger}$ (as can be checked exploiting that $\{\sigma_1, \sigma_2\} = 0$).  \\

Applying Eq. \eqref{defbell}, we can now get the explicit form of  $\hat{O}^{(max)}_{\rm Bell}$.

By direct calculation, we checked that
 \beq
 \langle \psi |\hat{O}^{(max)}_{\rm Bell}|\psi \rangle=  2.64575 \, ,
 \label{fval2}
 \eeq
 as reported in the main text.

The diagonalizing unitary matrix for $\hat{O}^{(max)}_{\rm Bell} = U^{\dagger} \, \hat{O}^{(max) \,  \mathrm{(diag)}}_{\rm Bell } \, U $  
can be applied to the initial state in Eq. \eqref{starg32}, in order to measure the CHSH inequality violation in a simpler way.
Indeed, 
\beq
\langle \psi | \hat{O}^{(max)}_{\rm Bell} | \psi \rangle = \langle \psi | U^{\dagger} \, \hat{O}^{(max) \,  \mathrm{(diag)}}_{\rm Bell } \, U | \psi \rangle = \langle \psi \, U^{\dagger} | \hat{O}^{(max)   \, \mathrm{(diag)}}_{\rm Bell } | U \,  \psi \rangle \, .
\eeq
Therefore, we can measure the diagonal operator $\hat{O}^{(max) \,  \mathrm{(diag)}}_{\rm Bell }$  \emph{in the diagonal $\big( \sigma_z,  S_z \big)$ basis}, provided that we measure it on the rotated state $\ket{\psi^{\prime}} = U \, \ket{\psi}$.  This is the approach followed in the main text.\\

Alternatively, one can measure, instead of the entire $\hat{O}^{(max)}_{\rm Bell}$, its components in Eq. \eqref{defbell}, for instance $A \otimes B$. We focus now on this operator, and on its matrix element:
\beq
\langle \psi | A \otimes B | \psi \rangle =\langle \psi | U_{AB}^{\dagger} \, \big( A \otimes B \big)^{(\mathrm{diag})} \, U_{AB}| \psi \rangle =\langle \psi | U_{AB}^{\dagger} \, P_1^{\dagger} \, \big( A^{(\mathrm{diag})} \otimes B^{(\mathrm{diag})} \big) \, P_1 \, U_{AB}| \psi \rangle\, ,
\label{sepU0}
\eeq
where 
\beq
P_1 \, U_{AB} =   \big(U_{A1} \otimes U_{B1}\big) \quad \Longleftrightarrow \quad  U_{AB} = P_1 \,  \big(U_{A1} \otimes U_{B1}\big) \ ,
\label{sepU}
\eeq 
the matrices $U_{A1}$ and $U_{B1}$ being defined in Eqs. \eqref{u31}, and
\beq
P_1 = P_1^{\dagger} = P_1^{-1} =\left(
\begin{array}{cccccccc}
 1 & 0 & 0 & 0 & 0 & 0 & 0 & 0 \\
 0 & 1 & 0 & 0 & 0 & 0 & 0 & 0 \\
 0 & 0 & 0 & 0 & 0 & 0 & 1 & 0 \\
 0 & 0 & 0 & 0 & 0 & 0 & 0 & 1 \\
 0 & 0 & 0 & 0 & 0 & 1 & 0 & 0 \\
 0 & 0 & 0 & 0 & 1 & 0& 0 & 0 \\
 0 & 0 & 1 & 0 & 0 & 0 & 0 & 0 \\
 0 & 0 & 0 & 1 & 0 & 0 & 0 & 0 \\
\end{array}
\right)
\eeq
being a unitary permutation.

Hence,
\beq
\ket{\psi^{\prime}_{AB}} =  U_{AB}  \, \ket{\psi} = P_1 \, \big( U_{A1} \otimes U_{B1}  \big) \, \ket{\psi} = P_1 \ket{\psi^{''}_{AB}} \, 
\eeq
\big($ \ket{\psi^{''}_{AB}} =   \big( U_{A1} \otimes U_{B1}  \big) \, \ket{\psi} $\big), and, also from Eqs. \eqref{defD} and \eqref{Adiags},
\beq
\big( A \otimes B \big)^{(\mathrm{diag})}=  P_1 \, \big( A^{(\mathrm{diag})} \otimes B^{(\mathrm{diag})} \big) \, P_1  =  P_1 \, \big(- \sigma_3 \otimes  D_1 \big) \, P_1\, .
\eeq
Therefore, we can conclude that
\beq
\langle \psi | A \otimes B | \psi \rangle = \langle \psi|  \big( U_{A1}^{\dagger} \otimes U_{B1}^{\dagger}\big)\big(- \sigma_3 \otimes  D_1 \big) \big( U_{A1} \otimes U_{B1}  \big)|\psi \rangle.
\label{sepU01}
\eeq
For our purposes, it is indeed convenient to measure the diagonal operator $\big(- \sigma_3 \otimes  D_1 \big)$ on the state $ \ket{\psi^{''}_{AB}} =   \big( U_{A1} \otimes U_{B1}  \big) \, \ket{\psi} $.

Notice also that, via the further (unitary) permutation of the diagonal basis,
\beq
P_2 = P_2^{\dagger} = P_2^{-1} = \left(
\begin{array}{cccccccc}
 1 & 0 & 0 & 0 & 0 & 0 & 0 & 0 \\
 0 & 0 & 0 & 0 & 0 & 1 & 0 & 0 \\
 0 & 0 & 0 & 0 & 0 & 0 & 1 & 0 \\
 0 & 0 & 0 & 0 & 0 & 0 & 0 & 1 \\
 0 & 0 & 0 & 0 & 1 & 0 & 0 & 0 \\
 0 & 1 & 0 & 0 & 0 & 0 & 0 & 0 \\
 0 & 0 & 1 & 0 & 0 & 0 & 0 & 0 \\
 0 & 0 & 0 & 1 & 0 & 0 & 0 & 0 \\
\end{array}
\right) \, , 
\eeq
 the even simpler (factorized) form
can be also obtained for the rotated diagonal operator
\beq
O_{AB}^{(\mathrm{diag})}  =  \sigma_3 \otimes  {\bf 1}_{4 \times 4} \, ,
\label{opdiagfin}
\eeq
to be measured on the state 
\beq
\ket{\psi^{'''}_{AB}} = P_2 \, \ket{\psi^{''}_{AB}} = P_2 \, \big( U_{A1} \otimes U_{B1}  \big) \, \ket{\psi} \, .
\eeq

Similar expressions can be obtained for the other addenda of  $\hat{O}^{(max)}_{\rm Bell}$ in Eq. \eqref{defbell}.

This second approach closely resembles that adopted for the $\big(\frac{3}{2} - \frac{3}{2} \big)$ case, where the unitary transformations $U^{(A)}$ and $U^{(B)}$ are applied.

\section{Measurements of the electronic and nuclear coherence times of Yb(trensal)}

\subsection{Synthesis}
Acetonitrile, tris(2-aminoethyl)amine, salicylaldehyde, Lu$_2$O$_3$, $^{173}$Yb$_2$O$_3$, and triflic acid were purchased from commercial sources and used as received. Lu(OTf)$_3\cdot9$(H$_2$O) was synthesized from Lu$_2$O$_3$ following a literature procedure \cite{Vesborg}.\\

{\bf $^{173}$Yb(OTf)$_3\cdot9$(H$_2$O)}: 
$^{173}$Yb$_2$O$_3$ (0.100 g, 0.25 mmol) was dispersed in 5 ml H$_2$O to which aqueous triflic acid (2.65 ml of 0.5653 M aq. solution, 1.5 mmol) was added. The mixture was refluxed overnight resulting in a clear solution. The solution was then filtered and left for slow evaporation yielding crystals of $^{173}$Yb(OTf)$_3\cdot9$(H$_2$O) (yield: 0.273 g, 70\%). \\

{\bf $^{173}$Yb$_0.0005$Lu$_99.9995$(trensal)}: Single crystals of $^{173}$Yb$_0.0005$Lu$_99.9995$(trensal) were prepared analogously to a literature procedure \cite{Pedersen2014}, employing a molar ratio of 0.05:99.95 for $^{173}$Yb(OTf)$_3\cdot9$(H$_2$O) and Lu(OTf)$_3\cdot9$(H$_2$O), respectively.

\subsection{Electron Paramagnetic Resonance}
Electron Paramagnetic Resonance (EPR) $T_{2e}$ relaxation times were measured on a single crystal of isotopically enriched $^{173}$Yb(trensal) doped at a concentration of 0.05$\%$ into a diamagnetic host of Lu(trensal). Measurements were performed on a commercial Bruker ELEXSYS-E580 spectrometer utilizing a ER 4118X-MD5 flexline resonator. The sample was cooled by flowing helium in an Oxford CF935O cryostat. $T_{2e}$ measurements were taken at ca. 9.7GHz microwave frequency utilizing the high-field line at 2018 G. Sample was oriented with the C$_3$ axis parallel to the main magnetic field z axis. Microwave $\frac{\pi}{2}$ pulse was set to either 20 ns or 60 ns.

\subsection{Nuclear Magnetic Resonance}
\begin{figure}[!h]
    \centering
\includegraphics[width=\textwidth]{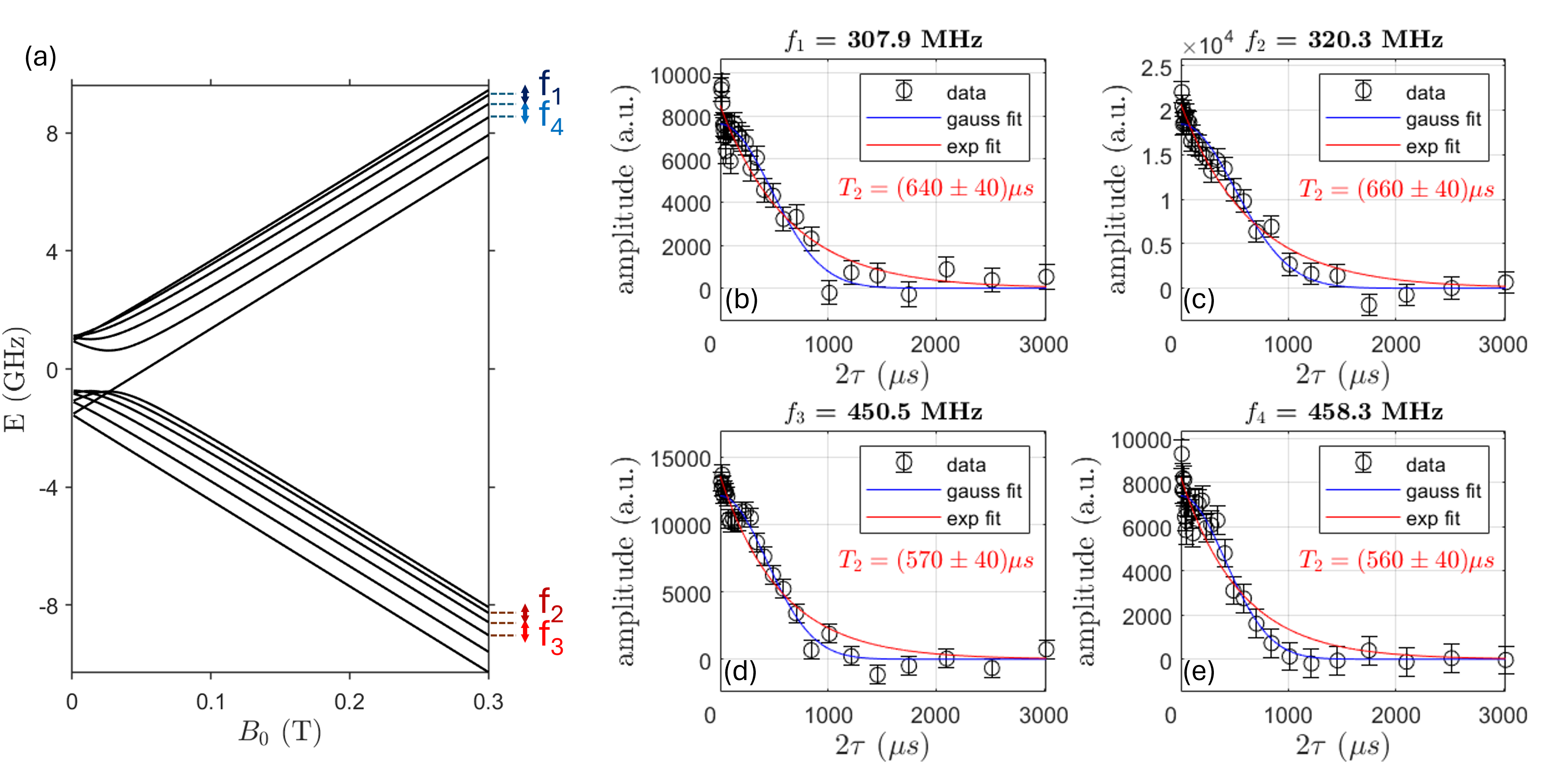}
    \caption{(a) Level diagram as a function of the external magnetic field applied along $z$, with the four transitions addressed by NMR indicated on the right. (b-e)    
    Hahn echo decay, for transitions $\ket{\uparrow,-\frac{1}{2}}\rightarrow\ket{\uparrow,-\frac{3}{2}}$ ($f_1$ = 307.9 MHz), $\ket{\downarrow,\frac{1}{2}}\rightarrow\ket{\downarrow,\frac{3}{2}}$ ($f_2$ = 320.3 MHz), $\ket{\downarrow,-\frac{1}{2}}\rightarrow\ket{\downarrow,\frac{1}{2}}$ ($f_3$ = 450.5 MHz) and $\ket{\uparrow,\frac{1}{2}}\rightarrow\ket{\uparrow,-\frac{1}{2}}$ ($f_4$ = 458.3 MHz) in a static magnetic field $B_z=0.3$ T applied along the $z$-axis at the base temperature of $T=1.4$ K. For each transition the amplitude of the signal as a function of the delay between the first pulse and the detected echo ($2 \tau$) is reported. Error bars  are obtained from the root-mean-square of the background noise (analyzed as signal). The red line shows the exponential fit, with the corresponding $T_2$ reported aside. The shortest $T_2$ (560 $\mu s$) was measured for $f_4$ and was used in the simulations reported in the main text. A gaussian fit (blue line) is also reported.}
    \label{fig:NMR_t2}
\end{figure}

The coherence times for the nuclear qudit were measured with Nuclear Magnetic Resonance (NMR) on a single crystal containing isotopically enriched $^{173}$Yb(trensal), doped at 0.05$\%$ into its diamagnetic Lu(trensal) isostructural analogue. The NMR experiments were performed with a home-built broadband NMR spectrometer optimized for magnetism, named ‘HyReSpect’ \cite{Hyrespect}, in a Maglab EXA (Oxford Instruments) superconducting magnet equipped with a helium flow variable temperature insert. The NMR probehead, consisting in a lumped LC resonant circuit with an interchangeable sample coil, yields a typical Q-factor in the order of 200 at low temperature and provides a tuning range of more than one octave. \\
We measured the coherence time $T_{2n}$ for transitions $\ket{\uparrow,-\frac{1}{2}}\rightarrow\ket{\uparrow,-\frac{3}{2}}$ ($f_1$ = 307.9 MHz), $\ket{\downarrow,\frac{1}{2}}\rightarrow\ket{\downarrow,\frac{3}{2}}$ ($f_2$ = 320.3 MHz), $\ket{\downarrow,-\frac{1}{2}}\rightarrow\ket{\downarrow,\frac{1}{2}}$ ($f_3$ = 450.5 MHz) and $\ket{\uparrow,\frac{1}{2}}\rightarrow\ket{\uparrow,-\frac{1}{2}}$ ($f_4$ = 458.3 MHz) in a static magnetic field $B_z=0.3$ T applied along the C$_3$ axis of the molecule at the base temperature of $T=1.4$ K.
We exploited a $\frac{2\pi}{3}-\tau-\frac{2\pi}{3}$ sequence \cite{Hahn_echo}, with two equal pulses of $500$ ns duration, a maximum delay $\tau = 1510$ $\mu$s and a radiofrequency power optimized for each transition to obtain the maximum signal. The detected echo amplitude is proportional to the component of the transverse magnetization precessing at the resonance frequency of the targeted transitions, and exhibits a decay as a function of the delay $\tau$ (see Fig.\ref{fig:NMR_t2}). The coherence decay for the four transition was fitted to the exponential function $M(\tau) = M_0e^{-2\tau/T_{2n}}$ with reasonable accuracy, with $T_{2n}$ values ranging from 560 $\mu$s to 660 $\mu$s. 
An exponential coherence decay is in fact consistent with the Lindblad master equation exploited in the numerical simulations described in the main text, 
although the coherence decay in such a magnetically diluted sample tends to a gaussian behavior. (This is typical of decoherence processes induced mainly by nuclear spins degrees of freedom).
In our simulations we considered the shortest $T_{2n}$ obtained from the fits in Figs.\ref{fig:NMR_t2}, in order to include the effect of pure dephasing in the less favorable conditions. \\

\section{Preparation of the $\big(\frac{1}{2}-\frac{3}{2}\big)$  states to probe  the CHSH inequality} 
\label{quditqubit}
\subsection{Realization of single and joint operations with resonant pulses}
To realize the quantum gates, we rely on resonant electromagnetic pulses acting on the coupled qubit-qudit system: {\color{black}Suitable sequences of electromagnetic pulses are exploited to implement single-qubit, single-qudit, and joint qubit-qudit transformations \cite{review}. These operations are based on transformations of the form
\begin{equation}
U_{x,y}(\theta, \phi) = \cos\Big(\frac{\theta}{2}\Big)(\ket{x}\bra{x} + \ket{y}\bra{y}) 
+ \sin\Big(\frac{\theta}{2}\Big)\left( \ket{y}\bra{x} e^{i\phi} - \ket{x}\bra{y} e^{-i\phi} \right)
+ \sum_{l\neq x,y} \ket{l}\bra{l},
\label{PRr}
\end{equation}
where $\ket{x}$ and $\ket{y}$ are eigenstates of the total system Hamiltonian that differ in one or both subsystems.
The same operations are realized via resonant square pulses of the form
\begin{equation}
\mathbf{B}_1(t) = B_1 \cos(\omega t + \phi)\,  \Theta(t - t_{\text{start}})\, \Theta(t_{\text{end}} - t)\,\hat{\mathbf{e}}_y,
\label{pulse}
\end{equation}
oriented along the $y$-axis. The frequency $\omega$ is tuned to the energy difference between $\ket{x}$ and $\ket{y}$, and the pulse duration defines the rotation angle $\theta$ \cite{tempiangolo}.\\
When the system features a constant (always-on) interaction between the qubit and the qudit, as in the present case, two main effects emerge. First, this interaction causes operations nominally addressing only one spin to act in a state-dependent way on the full qubit-qudit system.
As a result, pulses that are physically applied to a single degree of freedom can still generate entanglement between the qubit and the qudit. This enables the use of always-on interactions as a resource for quantum information processing, even in the absence of a tunable mediator. Designing such entangling gates, however, requires careful calibration of pulse parameters to account for the full dynamics of the coupled system.
Second, the states of each subsystem depend on the state of the other, so that implementing a single-qubit/qudit rotation within a given subsystem requires a set of pulses for each possible state of the other.}\\
An important parameter determining the fidelity of the implemented rotation is the pulse amplitude
$B_1$. In principle, smaller $B_1$ better allow for higher frequency selectivity and hence lower the leakage. However this leads to longer pulses, thus enhancing the effect of decoherence. The optimal strategy is based on a trade-off between these two effects and depends on the value of $T_{2e}$. 

\section{Preparation of the entangled (1/2-3/2) state}
Starting from the state  $\ket{-\frac{1}{2}, -\frac{5}{2}}$, we apply a sequence of five pulses to reach the state 
\beq
\ket{\psi} = \frac{1}{2} \, \Bigg(\ket{\frac{1}{2},\frac{3}{2}} + \ket{\frac{1}{2},-\frac{1}{2}} + \ket{-\frac{1}{2},-\frac{1}{2}} + \ket{-\frac{1}{2},-\frac{3}{2}} \Bigg) \, .
\label{starg32}
\eeq
In particular, this sequence is composed by: 
\begin{enumerate}
    \item  a radiofrequency transition between the qudit states $\ket{x} = \ket{-\frac{5}{2}}$ and $\ket{y} = \ket{-\frac{3}{2}}$, with $\theta = \pi$ and $ \phi = \pi$ (referred to the expression $U_{xy}(\theta, \phi)$ in Eq. \eqref{PRr}); 
    \item a transition between the qudit states $\ket{-\frac{3}{2}}$ and $\ket{-\frac{1}{2}}$, with parameters $ \theta = \frac{2 \, \pi}{3}$ and $\phi = \pi$;
    \item a microwave transition between the qubit states $\ket{-\frac{1}{2}}$ and $\ket{\frac{1}{2}}$, with parameters $ \theta = 2 \, \arcsin \sqrt{\frac{2}{3}}$ and $\phi = \pi$. Due to the higher transition frequency of the electronic spin, the microwave pulse is significantly faster than the radiofrequency pulses acting on the nuclear qudit. In particular, its duration ranges between 7 and 14 ns depending on the assumed amplitude $B_1$ of the oscillating field. In comparison, the full sequence of five operations requires a total amount of time between 345 and 1704 ns; 
    \item a transition between the qudit states $\ket{-\frac{1}{2}}$ and $\ket{\frac{1}{2}}$, with parameters $ \theta = \frac{\pi}{2}$ and $\phi = 0$;
    \item a transition between the qudit states $\ket{\frac{1}{2}}$ and $\ket{\frac{3}{2}}$, with parameters $\theta = \pi$ and $\phi = 0$.
\end{enumerate}
The results obtained from this procedure are reported in Table \ref{prova}: the optimization of the pulse amplitudes has been performed manually. The amplitudes of the  pulses described above, along with the fidelity values with and without decoherence, are summarized in the same table. 
\begin{table}[!ht]
    \centering
    \begin{tabular}{llllllllllllll}
    \hline
      $B_{1}$ (G) &$\mathcal{F}$ $(T_{2e} = 20 \, \mu$s) & $\mathcal{F}$ $(T_{2e} = 10 \, \mu$s) &$\mathcal{F}$ $(T_{2e} = 5 \, \mu$s) &  $\mathcal{F}$ $(T_{2e} = 3 \, \mu$s) & $\mathcal{F}$ $(T_{2e} = 2.4 \, \mu$s) &  $\mathcal{F}$ $(T_{2e} = 2 \, \mu$s) & $\mathcal{F}$ $(T_{2e} = 1 \, \mu$s) & Time ($\mu$s) \\   \hline
       10   & 0.9770 & 0.9600 & 0.9279 & 0.8884 & 0.8956 & 0.8441 & 0.7392 & 1.7720 \\
      15   & 0.9770 & 0.9656 & 0.9435 & 0.9156& 0.8991 & 0.8832 & 0.8002 & 1.1813 \\
      20  & 0.9809 & 0.9722 & 0.9551 & 0.9333 & 0.9202 & 0.9075 & 0.8389 & 0.8942 \\
      25  & 0.9877 & 0.9806 & 0.9668 & 0.9490 & 0.9383 & 0.9278 & 0.8698 & 0.7144 \\
       30  & 0.9818 & 0.9759 & 0.9644 & 0.9495 & 0.9404 & 0.9315 & 0.8816 & 0.5962  \\
        40 & 0.9610 & 0.9567 & 0.9482 & 0.9370 & 0.9302 & 0.9235 & 0.8853 & 0.4471 \\
         60  & 0.9221 & 0.9218 & 0.9163 & 0.9091 & 0.9046 & 0.9002 & 0.8746 &  0.2981\\\hline
      \end{tabular}
      \caption{Fidelity $\mathscr{F}$ with the state in Eq. \eqref{starg32} for different values of $B_{1}$, with and without decoherence. The time required to reach the target state for the different amplitude values is also reported.}
 \label{prova}
\end{table}

\section{Implementation of the unitary operators to measure $O_{Bell}$}
To overcome excessive leakage effects, pulse shaping techniques have been employed, in particular the Gradient Ascent Pulse Engineering (GRAPE) optimization algorithm  \cite{nori2012,nori2013} from QuTiP \cite{lambert2024qutip5quantumtoolbox}. This approach effectively suppressed leakage, enabling the realization of precise operations in significantly shorter times compared to the square-shape approach exploited so far.\\
More in detail, the GRAPE algorithm designs optimal control pulses for quantum systems by discretizing control fields into \( N \) time segments, each with constant amplitudes \(\{c_k^{(j)}\}\), $j$ labelling the time intervals and $k$ the different pulses to optimize at the different drivings to optimize in parallel. Consequently, the dynamics in the $j$-th interval is governed by the time-dependent Schröedinger equation:  
\begin{equation}  
i\hbar\frac{d}{dt}|\psi(t)\rangle = H_j(t)|\psi(t)\rangle, \quad H_j(t) = H_0 + \sum_{k=1}^M c_k^{j}(t)H_k \, .  
\end{equation}  
The total evolution operator \( U(T) \) is computed as a time-ordered product of  unitaries  
\begin{align}  
U(T) &= \mathcal{T} \prod_{j=1}^N U_j \, , \quad U_j = e^{-iH_j\Delta t/\hbar} \, , \\  
H_j &= H_0 + \sum_{k=1}^M c_k^{(j)}H_k \, ,  
\end{align}  
where \( \mathcal{T} \) enforces time ordering. By defining $U_{target}=U_{Ai}\otimes U_{Bj}$ (defined in in  Eq. \eqref{defB1a}, Eq. \eqref{defB2a}, and Eq. \eqref{u31}), the analytic gradient of the fidelity \(\mathcal{F} = |\langle U_{\text{target}}|U(T)\rangle|^2\) is computed via the quantum chain rule, enabling gradient ascent maximization. This procedure dynamically reshapes \( H(t) \) to achieve high-fidelity quantum operations, balancing computational efficiency and experimental robustness \cite{PhysRevA.69.022319, de_Fouquieres_2011, glaser_2015}. 

\begin{figure}[!t]
    \centering
    \includegraphics[width=0.6\textwidth]{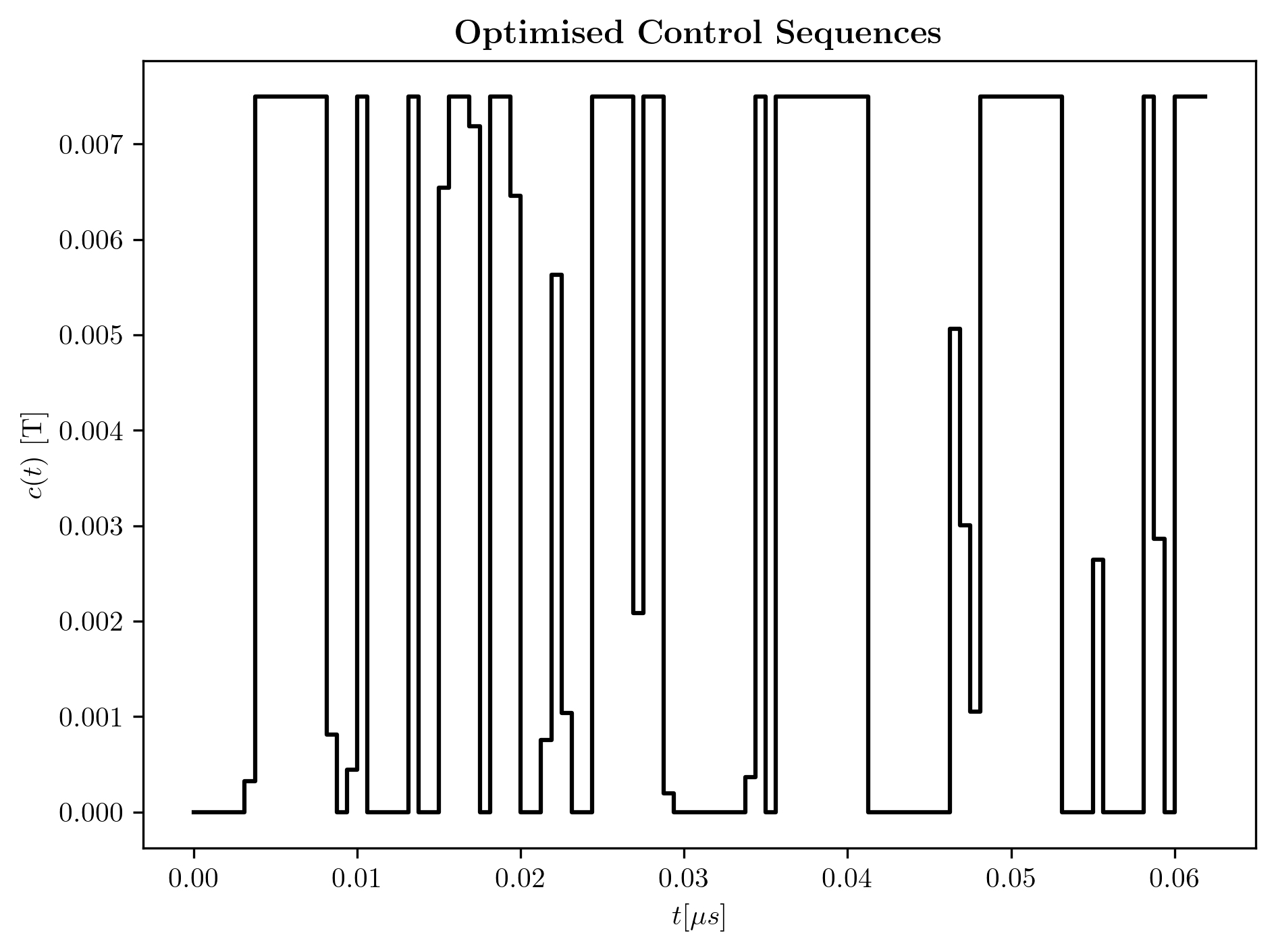}
    \caption{
Optimized control sequence $c(t)$ the driving Hamiltonian $H_d$, which represents the magnetic moment along the $z$-axis.  The plot shows the first 100 points of the 1600-point sequence, corresponding to a time interval of $t\in[0,0.0625]\mu$s.  The total sequence duration is $1\mu s$, discretized into 1600 time steps to limit the control bandwidth. The control pulse is unique, as only one driving Hamiltonian is considered in the system.}
    \label{grapepulse}
\end{figure}

In the square-shape approach, leakage reduction required lowering the pulse amplitude; in this case, this reduction would make the procedure too slow and susceptible to decoherence. In contrast, by leveraging GRAPE and pulse shaping techniques, the optimized pulses achieved the desired target states in 1 $\mu$s, with a maximum pulse amplitude of 75 Gauss and frequency components not exceeding 800 MHz. These values are currently achievable in broadband multi-frequency NMR experiments with molecular nanomagnets. \\
We remark that the GRAPE approach not only improved the precision of the gate operations but also minimized the impact of decoherence by making operations significantly faster.\\
A part of one of the four GRAPE-generated control pulses which is designed to implement one of the target unitaries is shown in Fig. \ref{grapepulse}.
The optimized control pulses were employed to simulate the system evolution, considering both coherent dynamics and the effects of decoherence. The expectation values of $\hat{O}^{max}_{\rm Bell}$, calculated from the states listed in Table \ref{prova}, are presented in Table \ref{omax}. {\color{black}The fidelities of the states obtained from the entire procedure with the target states are summarized for the most favourable cases in Table \ref{grape}. There we also report a comparison between the results obtained using the GRAPE pulses and the best results from hand-optimized sequences of pulses, modeled as in Eq. \eqref{pulse} of the main text.\\
Finally, to better assess the effectiveness of the two approaches, in Table \ref{grapecompare} we compare the values of $\hat{O}^{max}_{\rm Bell}$ obtained in the best-case scenarios for each decoherence time \(T_{2e}\), both for GRAPE pulses and for hand-optimized sequences.
\begin{table}[!ht]
    \centering
     \begin{tabular}{lllllllllllll}
    \hline
      $B_{1}$ (G)  
       &   $(T_{2e} = 20 \, \mu$s)
       &    $(T_{2e} = 10 \, \mu$s) &    $(T_{2e} = 5 \, \mu$s) &   $(T_{2e} = 3 \, \mu$s)  &  $(T_{2e} = 2.4 \, \mu$s)   &  $(T_{2e} = 2 \, \mu$s) &$(T_{2e} = 1 \, \mu$s)  \\   \hline
     15  & 2.4606 & 2.4327 & 2.3584 & 2.2689 & 2.2167 & 2.1673 & 1.9253 \\
      20  & 2.4872 & 2.4534 & 2.3857 & 2.3019 & 2.2537  & 2.2074 & 1.9758\\
      25  & 2.5630 & 2.5299 & 2.4664 & 2.3876 & 2.3413 & 2.2972 & 2.0732 \\
      30  & 2.5392 & 2.5090 & 2.4523 & 2.3788 & 2.3362 & 2.2960 & 2.0878 \\
      40  & 2.4675 & 2.4396 & 2.4109 & 2.3181 & 2.2785 & 2.2403 & 2.0427 \\\hline
      \end{tabular}
      \caption{Expectation values of $\hat{O}^{max}_{\rm Bell}$, calculated starting from the states obtained by using an amplitude $B_1$ listed in Table \ref{prova} for various dephasing times.}
 \label{omax}
\end{table}
\begin{table}[!ht]
\centering
\begin{tabular}{llllllllll}
\hline
 & \multicolumn{4}{c}{GRAPE pulses} & \multicolumn{4}{c}{Hand-optimized pulses} \\

 & $T_{2e} = 10 \, \mu$s & $T_{2e} = 5 \, \mu$s & $T_{2e} = 3 \, \mu$s &$T_{2e} = 2.4\, \mu$s  & $T_{2e} = 10 \, \mu$s & $T_{2e} = 5 \, \mu$s & $T_{2e} = 3 \, \mu$s &$T_{2e} = 2.4 \, \mu$s\\
\hline
$\rho_{A1,B1}$ & 0.9407 & 0.9089 & 0.8695 & 0.8464 & 0.7906 & 0.6935 & 0.6081 & 0.5904 \\
$\rho_{A1,B2}$  & 0.9409 & 0.9182 &  0.8819 & 0.8606  & 0.7891 & 0.6957 & 0.6046 & 0.6515 \\
$\rho_{A2,B1}$  & 0.9374 &  0.9061 & 0.8634 & 0.8387 & 0.7768 & 0.6853 & 0.6609 & 0.5824 \\
$\rho_{A2,B2}$  & 0.9360  &  0.9051 & 0.8650 & 0.8416 & 0.7834 & 0.6674& 0.6674 & 0.6364 \\
\hline
\end{tabular}
      \caption{Fidelity $\mathcal{F}$ between the ideal target state — obtained by applying the unitaries in Eq.~\eqref{u31}, Eq.~\eqref{defB1a}, and Eq. \eqref{defB2a} to the state in Eq.~\eqref{starg32} — and the best-case results from our complete simulation procedures. In particular, when using the GRAPE algorithm, the simulation combines the  preparation of the state in Eq. \eqref{starg32}, via hand-optimized pulses, with GRAPE-optimized unitary operations. Initial entangled states were prepared with varying $B_1$, for each pulse: without decoherence ($B_1 = 20$, $\mathcal{F} = 0.9907$), and with decoherence times $T_{2e} = 10\mu$s ($B_1 = 20$, $\mathcal{F} = 0.9724$), $T_{2e} = 5\mu$s ($B_1 = 20$, $\mathcal{F} = 0.9563$), and $T_{2e} = 3\mu$s ($B_1 = 30$, $\mathcal{F} = 0.9514$). After reaching the entangled state, the GRAPE algorithm optimized the pulses to minimize leakage and decoherence effects. The hand-optimized  sequences, instead, were built starting from the same initial entangled states used for GRAPE and manually tuning the amplitudes in Eq.\eqref{pulse} to find a trade-off between leakage and decoherence, as discussed in the main text. The reported fidelities characterize the final states as result of the entire process.}
 \label{grape}
\end{table}
\begin{table}[h!]
\centering
\begin{tabular}{lcc}
\hline
\textbf{\(T_{2e}\)} & $\hat{O}^{max}_{\rm Bell}$ (GRAPE) &$\hat{O}^{max}_{\rm Bell}$ (hand-optimized) \\
\hline
10\(\mu\)s     & 2.5299 &  1.7744\\
5\(\mu\)s        &  2.4664 & 1.4595  \\
3\(\mu\)s         & 2.3876 &  1.3578 \\
2.4\(\mu\)s         & 2.3413  & 1.3178 \\
\hline
\end{tabular}
\caption{Comparison of the best values of $\hat{O}^{max}_{\rm Bell}$ obtained with GRAPE and by hand-optimized pulse sequences, for different decoherence times \(T_{2e}\).}
\label{grapecompare}
\end{table}}
\section{Preparation of the target $\big(\frac{3}{2} - \frac{3}{2}\big)$-states}
\label{apprep}

\subsection{Realization of two-qudit gates via ancilla-controlled interactions}
{\color{black} To control the \big($\frac{3}{2}$-$\frac{3}{2}$\big) system,  we use both single-qudit rotations, that are implemented as for the \big($\frac{1}{2}$-$\frac{3}{2}$\big), and two-qudit controlled operations, that, due to the absence of direct interaction between the two qudits, requires the presence of a switchable interaction \cite{switch3,Switch, PhysRevResearch.4.043135,switch4, FERRANDOSORIA2016727} controlled by a selectively-excited ancilla qubit.  In particular, we implement a two-qudit controlled-$Z$ gate \cite{review} defined as
\begin{align}
W_{\mu, \nu}(\pi) = e^{i\pi} \ket{\mu\nu}\bra{\mu\nu} + \sum_{lj \neq \mu\nu} \ket{lj}\bra{lj},
\end{align}
which applies a phase $\pi$ to the specific component $\ket{\mu\nu}$ of the two-qudit wavefunction.\\
This transformation is realized by applying a selective excitation and de-excitation of the ancilla, conditioned on the joint state $\ket{\mu\nu}$ of the two qudits. That is, the ancilla undergoes a $\pi$-rotation only if the two-qudit subsystem is in the state $\ket{\mu\nu}$. As a result, a relative phase of $\pi$ is induced solely on the $\ket{\mu\nu}$ component, effectively implementing the controlled-$Z$ operation.\\
Crucially, unlike in the $\big(\frac{1}{2}-\frac{3}{2}\big)$ system, the interaction is switchable: it is only active during the gate operation and does not affect the dynamics of the qudits afterward \cite{Switch}, thereby preserving their independence after entanglement. This holds strictly to first order, apart from possible residual couplings.\\
\subsection{Preparation of the maximally entangled state}
Starting from the state
\begin{equation}
    \ket{\psi}=\ket{-\frac{3}{2}}_1\otimes \ket{-\frac{1}{2}}_2 \otimes \ket{-\frac{3}{2}}_3 \, ,
    \label{eq1a}
\end{equation}
to reach the maximally entangled state 

\begin{align}
& \ket{\psi}=\frac{1}{2}(\ket{-\frac{3}{2},-\frac{1}{2}, -\frac{3}{2}}+\ket{-\frac{1}{2},-\frac{1}{2}, -\frac{1}{2}}+\ket{+\frac{1}{2},-\frac{1}{2}, +\frac{1}{2}}
+\ket{+\frac{3}{2},-\frac{1}{2}, +\frac{3}{2}})=  \label{eq42}\\ 
&\frac{1}{2}(\ket{-\frac{3}{2}, -\frac{3}{2}}+\ket{-\frac{1}{2}, -\frac{1}{2}}+\ket{+\frac{1}{2}, +\frac{1}{2}}
+\ket{+\frac{3}{2}, +\frac{3}{2}})_{13}\otimes\ket{-\frac{1}{2}}_2 \, ,  \nonumber
\end{align}

we implement a sequence of 15 pulses.
In particular, we first act on the first qudit with the transformations
\begin{equation}
    U_{1,2}^{(1)}U_{2,3}^{(1)}U_{3,4}^{(1)} \, ,
    \label{un1}
\end{equation} 
obtaining the state
\beq
\ket{\psi}=\frac{1}{2}(\ket{-\frac{3}{2}}+\ket{-\frac{1}{2}}+\ket{+\frac{1}{2}}+\ket{+\frac{3}{2}})_1\otimes\ket{-\frac{1}{2}}_2 \otimes \ket{-\frac{3}{2}}_3
\label{cohe}
\eeq
Specifically, we defined $\ket{1} \equiv \ket{+\frac{3}{2}}$, $\ket{2}\equiv\ket{+\frac{1}{2}}$, $\ket{3} \equiv \ket{-\frac{1}{2}}$ and $\ket{4} \equiv\ket{-\frac{3}{2}}$; the superscript indicates the qudit  which the rotations in Eq. \eqref{un1} acts on.
Table \ref{tabro} reports the optimal parameters for the electromagnetic pulses used to implement the rotations in Eq. \eqref{un1}, considering the need to balance pulse precision and the presence of decoherence. 
The pulses are divided in two  groups based on the transitions they address:
\begin{enumerate}
\item for the $\frac{\pi}{2}$ single-qudit rotations, a larger amplitude (e.g., 70 G) is chosen. Indeed, due to the residual qudit-qudit interaction still present when the switch is off, for this system the transition frequencies for single-qudit operations can be slightly influenced by the state of the other qudit. The error associated to this effect can be mitigated by using broader frequency pulses (i.e. by increasing $B_1$ to reduce $t_{end}-t_{start}$).\\
\item in contrast, for all other pulses a smaller amplitude (e.g., 20 G) is used. For two-qudit operations, the transitions are naturally well-targeted.
 A smaller amplitude is also used for certain single-qudit transitions. This is possible because, when applying the gate, the state of the entire system is known and the state of the unmodified qudit does not influence the energy of the single-qudit transition. 
\end{enumerate}
By adopting this differentiated approach, the pulse sequences are optimized to address the distinct challenges of broad-spectrum operations versus high-precision transitions, ensuring robust performance even in the presence of decoherence.

\begin{table}[h]
\begin{tabular}{llllll}
\hline
\multicolumn{5}{c}{First Qudit Rotations}                                                                    \\
 Initial State & Final State &  $\theta$         &  $\phi$ & Optimal $B_1$ (G) & $\omega$ (ns$^{-1}$)\\ \hline
$\ket{-\frac{3}{2}}$           & $\ket{-\frac{1}{2}}$        & $\frac{2\pi}{3}$                   & $\pi$    & 30       & 315.82                       \\
 $\ket{-\frac{1}{2}}$             & $\ket{+\frac{1}{2}}$         & 2 arcsin$\sqrt{\frac{2}{3}})$ & $\pi$     & 30    & 224.95                         \\
$\ket{+\frac{1}{2}}$             & $\ket{+\frac{3}{2}}$           & $\frac{\pi}{2} $                  & $\pi$    & 30     & 133.82                        \\ \hline
\end{tabular}
\caption{Parameters for the three single-qudit rotations in Eq. \eqref{un1} performed on the first qudit. The pulses are the ones described in Eq. \eqref{pulse}, where $\hbar \omega$ is the difference between the energies of the final and of the initial states.}
\label{tabro}
\end{table}

At this point, 12 further gates are executed, including nine single-qudit gates on the second qudit and three two-qudit gates, in the following order:
\begin{itemize}
    \item we transform state in Eq. \ref{cohe} into the state 
    \begin{align}
& \frac{1}{2}(\ket{-\frac{3}{2}, -\frac{1}{2}, -\frac{3}{2}}+\ket{-\frac{1}{2}, -\frac{1}{2}, -\frac{3}{2}} \nonumber \\
&+\ket{+\frac{1}{2}, -\frac{1}{2}, -\frac{3}{2}}+\ket{+\frac{3}{2}, -\frac{1}{2}, +\frac{3}{2}}) 
    \end{align}
 applying the gates
    \begin{equation}
        U_{1,2}^{(2)}U_{2,3}^{(2)}U_{3,4}^{(2)}W_{13}^{(12)}U_{3,4}^{(2)}.
    \end{equation}
In particular, the sequence of three pulses $U_{3,4}^{(2)}W_{13}^{(12)}U_{3,4}^{(2)}$ allows for the implementation of an entangling CNOT gate \cite{switch4};
\item  we implement a CNOT gate again, utilizing the pulses 
\begin{equation}
    U_{2,3}^{(2)}U_{3,4}^{(2)}W_{23}^{(12)}U_{3,4}^{(2)}
\end{equation}
to reach the state
 \begin{align}
& \frac{1}{2}(\ket{-\frac{3}{2}, -\frac{1}{2}, -\frac{3}{2}}+\ket{-\frac{1}{2}, -\frac{1}{2}, -\frac{3}{2}} 
+\ket{+\frac{1}{2}, -\frac{1}{2}, +\frac{1}{2}}+\ket{+\frac{3}{2}, -\frac{1}{2}, +\frac{3}{2}});
\label{maxs}
    \end{align}
 \item   the state in Eq. \eqref{eq42} is eventually obtained with the pulses
\begin{equation}
    U_{3,4}^{(2)}W_{33}^{(12)}U_{3,4}^{(2)}. 
\end{equation} 
\end{itemize}
A detailed description of the parameters for these rotations is provided in  in Table \ref{tabro2}.\\
\begin{table}
\begin{tabular}{llllll}
\hline
\multicolumn{5}{c}{Second Qudit and Two-Qudit Rotations}                                                                    \\
 Initial State & Final State &  $\theta$         &  $\phi$ & Optimal $B_1$ (G) & $\omega$ (ns$^{-1})$\\ \hline
$\ket{-\frac{3}{2}}$           & $\ket{-\frac{1}{2}}$        & $\frac{\pi}{2}$                   & $\pi$     & 70         & 286.90                      \\
 $\ket{+\frac{3}{2}, -\frac{1}{2},-\frac{1}{2}}$             & $\ket{+\frac{3}{2}, +\frac{1}{2},-\frac{1}{2}}$         & $\pi$ & 0     & 30  & 479.21                           \\
$\ket{-\frac{3}{2}}$           & $\ket{-\frac{1}{2}}$        & $\frac{\pi}{2}$                     &   0   & 70           & 286.90                  \\ 
$\ket{-\frac{1}{2}}$           & $\ket{+\frac{1}{2}}$        & $\pi$                     &   0   & 30            & 198.81                 \\ 
$\ket{+\frac{1}{2}}$           & $\ket{+\frac{3}{2}}$        & $\pi$                     &   $\pi $ & 30                  & 159.95           \\
$\ket{-\frac{3}{2}}$           & $\ket{-\frac{1}{2}}$        & $\frac{\pi}{2}$                   & $\pi$     & 70     &          281.42                \\
 $\ket{+\frac{1}{2}, -\frac{1}{2},-\frac{1}{2}}$             & $\ket{+\frac{1}{2}, +\frac{1}{2},-\frac{1}{2}}$         & $\pi$ & 0     & 30                          & 493.73   \\
$\ket{-\frac{3}{2}}$           & $\ket{-\frac{1}{2}}$        & $\frac{\pi}{2}$                     &   0   & 70  & 281.42                           \\
$\ket{-\frac{1}{2}}$           & $\ket{+\frac{1}{2}}$        & $\pi$                     &   0   & 30   & 220.71  \\
$\ket{-\frac{3}{2}}$           & $\ket{-\frac{1}{2}}$        & $\frac{\pi}{2}$                   & $\pi$     & 70       & 281.42                       \\
 $\ket{-\frac{1}{2}, -\frac{1}{2},-\frac{1}{2}}$             & $\ket{-\frac{1}{2}, +\frac{1}{2},-\frac{1}{2}}$         & $\pi$ & 0     & 30    & 486.28                          \\
$\ket{-\frac{3}{2}}$           & $\ket{-\frac{1}{2}}$        & $\frac{\pi}{2}$                     &   0   & 70      & 281.42                       \\\hline
\end{tabular}
\caption{Parameters for the  single-qudit rotations performed on the second qudit and the two-qudit rotations.}
\label{tabro2}
\end{table}To achieve the best possible fidelity with the target state, various tests were performed by modifying the amplitudes $B_1$ of the electromagnetic pulses and for different values of the decoherence time $T_2$. 
The most significant results are shown in Table \ref{tab2}, where we also report the times required for the implementation  of the transformations as the pulse amplitudes vary.
Specifically, when $T_2$ is in the realistic range [10-30] $\mu s$, the best fidelities are obtained using an amplitude of 70 G for the $\frac{\pi}{2}$ single-qudit rotations (the first group mentioned just above),  and an amplitude of 30 G for the remaining transformations, that means the second group.
The results summarized in Table  \ref{tab2} are depicted graphically in Figure \ref{377}.\\\begin{table}[!ht]
    \centering
    \begin{tabular}{lllll}
    \hline
        $B_1$(G)  & $T_2=5 \mu s$ & $T_2=10  \mu s$& $T_2=30 \mu s$ & No decoherence \\ \hline 
       20 - 40 & 0.8973 & 0.9417 & 0.9739 & 0.9908 \\
       05 - 70 & 0.7265 & 0.8378 & 0.9350 & 0.9928 \\
       15 - 70 & 0.8790 & 0.9322 & 0.9714 & 0.9923 \\ 
       20 - 70 & 0.9011 & 0.9434 & 0.9739 & 0.9898 \\ 
       25 - 70  & 0.9098 &  0.9449 & 0.9699 & 0.9829   \\ 
       40 - 70 & 0.9162 & 0.9400 & 0.9565 & 0.9650 \\ 
       20 - 90 & 0.9011 & 0.9422 & 0.9717 & 0.9872 \\ \hline
    \end{tabular} 
    \hspace{1cm}
    \begin{tabular}{ll}
    \hline
        $B_1$(G) & Time(ns)  \\ \hline
        20 - 40 & 134.49 \\
        05 - 70 & 484.77 \\
        15 - 70 & 167.50 \\ 
        20 - 70 &  127.84 \\
        25 - 70 &  104.05 \\
        40 - 70 & 68.35 \\
        20 - 90 & 125.87 \\ \hline
    \end{tabular}
\caption{Left panel: fidelity $\mathscr{F}$ with the state in Eq. \eqref{eq42} for different values of $T_2$ and $B_1$. The two amplitude values correspond to the two different groups of pulses introduced  above. Right panel: total duration of the 15 transformations required to obtain the target state in Eq. \eqref{eq42}. Using smaller amplitudes of the pulses as in Eq. \eqref{pulse} results in significantly longer durations for the entire procedure, then larger dephasing effects.}
\label{tab2}
\end{table}
\begin{figure}[!ht]
    \centering
    \includegraphics[width=0.48\textwidth]{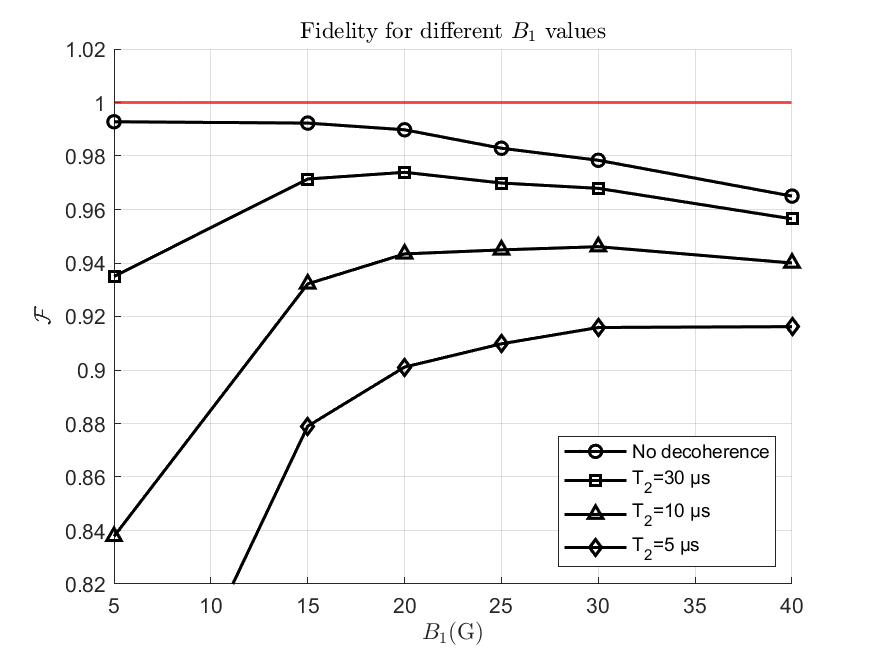}
    \includegraphics[width=0.48\textwidth]{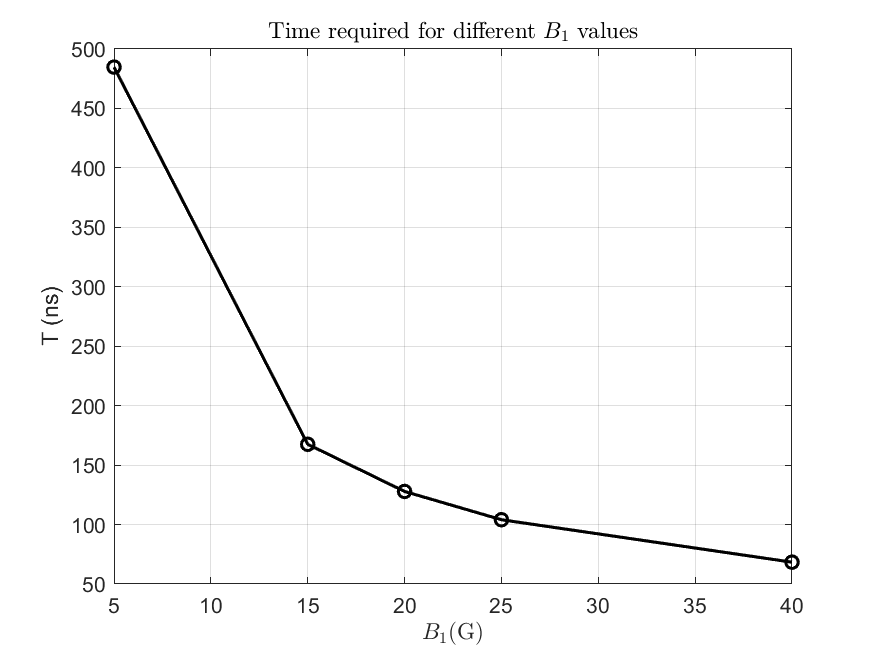}
    \caption{Summary of the results in Table  \ref{tab2}.  Left panel: on the $x$-axis, we report the amplitude values $B_1$ in Gauss for the pulses in the second group defined after Eq. \eqref{cohe}, setting the amplitudes for the first group equal to their optimal value 70 G. On the $y$-axis, we report the fidelity between  the ideal target state in Eq. \eqref{eq42} and its optically-synthesized counterpart. Right panel: we report the time required to perform the entire synthesizing process, varying the amplitudes $B_1$ of the pulses in the second group and setting the amplitudes in the first group all equal to 70 G.}
    \label{377}
\end{figure}
Eventually, Fig. \ref{figura3} shows a graphical representation of the temporal evolution of the nonvanishing diagonal elements of the density matrix of the desired final state in Eq. \eqref{eq42}.
\begin{figure}[h]
\centering
\includegraphics[width=0.5\textwidth]{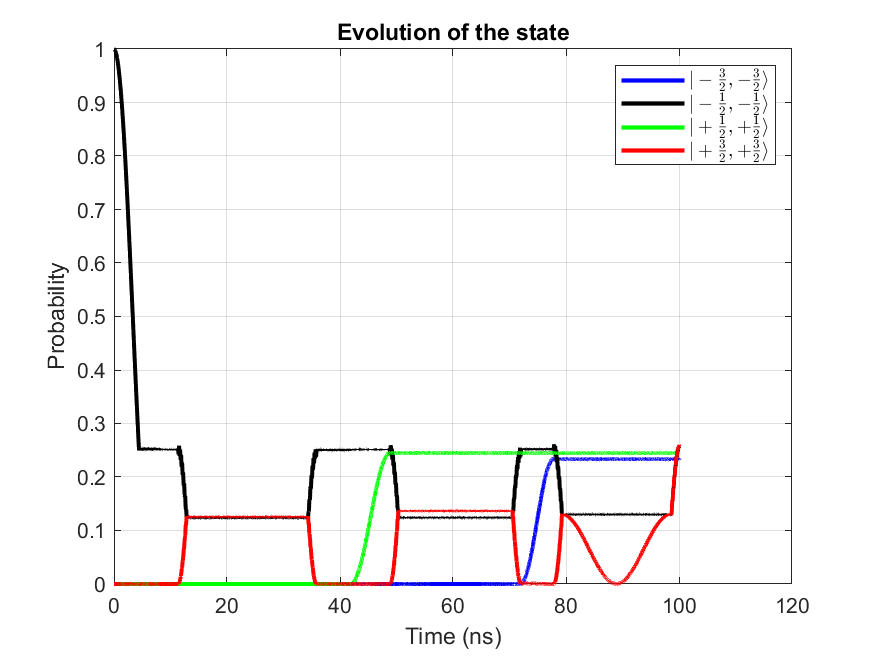}
\caption{Time evolution of the nonvanishing diagonal elements of the density matrix representing the state in Eq. \eqref{eq42}. These elements converge to almost equal values over time, achieving a maximally entangled state around 100 ns. Given that the noise primarily manifests as pure dephasing, any visible imperfection in the plot can be attributed to leakage effects.}
\label{figura3}
\end{figure}

\subsection{Implementation of the unitaries}
Once the maximally-entangled state in Eq. \eqref{maxs} is achieved, before performing the measurements to verify the violation of the inequality,
 we aim to apply the unitary operations with matrix elements \cite{Polozova}:
 \begin{equation}
[U^{(A)}_i]_{k,l}=\frac{1}{\sqrt{d}}e^{\frac{2\pi i \,  l}{d}(\alpha_i+k)} \, , \quad \quad
[U^{(B)}_j]_{k,l}=\frac{1}{\sqrt{d}}e^{\frac{2\pi i \, l}{d}(\beta_j-k)} \, ,
\label{eq9supp}
\end{equation}
where $d=4$, $\{i,j\}=1,2$ (not related to the spin-indices), and $\alpha_1=0$, $\alpha_2=\frac{1}{2}$, $\beta_1=\frac{1}{4}$, $\beta_2=-\frac{1}{4}$. 
 It is therefore necessary to convert these unitaries in electromagnetic pulses. \\
 For this task, we consider that an arbitrary transformation $W\in SU(d)$ can be decomposed in planar rotations using an iterative procedure, described e.g. in \cite{PhysRevResearch.4.043135, D'Alessandro:1251242}. First, we multiply
$W$ on the right-hand side by $U_{d-1,d}(\theta_{\frac{d(d-1)}{2}},\beta_{\frac{d(d-1)}{2}})$ so that the $(d,d-1)$ element of $W_1=WU_{d-1,d}\Big(\theta_{\frac{d(d-1)}{2}},\beta_{\frac{d(d-1)}{2}}\Big)$ is zero. Then, we multiply $W_1$ on the right-hand side by $U_{d-2,d}\Big(\theta_{\frac{d(d-1)}{2}-1},\beta_{\frac{d(d-1)}{2}-1}\Big)$ so that the $(d,d-2)$ element of $W_2=W_1\,U_{d-2,d}\Big(\theta_{\frac{d(d-1)}{2}-1},\beta_{\frac{d(d-1)}{2}-1}\Big)$ is zero. The procedure is repeated until we have the $W_{d-1}$ matrix that has all zero in the $d$ column and in the $d$ row, except for the $(d, d)$th element which must have magnitude 1 because the matrix is unitary. The same scheme could be repeated for all the other lines \cite{PhysRevResearch.4.043135}, without affecting further the $d$-th column and row. 
Following this approach, we conclude that for a four-by-four unitary matrix $W$, the product 
\begin{align}
&WU_{3,4}U_{2,4}U_{1,4}U_{2,3}U_{1,3}U_{1,2}=e^{i\lambda}P_{1,2}P_{2,3}P_{3,4} \, ,
\label{eq3333}
\end{align}
is diagonal;
$U_{x,y}$ are planar rotation defined in Eq. \eqref{PRr} and the matrices $P_{x,y}(\alpha)$, also diagonal, are defined as
\begin{equation} 
    P_{x,y}(\alpha)=\ket{x}\bra{x}e^{i\alpha}+\ket{y}\bra{y}e^{-i\alpha}+\sum_{l\neq x,y}\ket{l}\bra{l}.
 \label{defP}   
\end{equation}\\
Therefore, introducing the transitions $\pi_{x,y}^{\pm}$, which are $\pi$ planar rotations with phase either equal $\pi$ or zero, properly used to swap the levels in order to make the transitions between non-neighboring states possible, we can write the four-by-four unitaries  $U_a^{(x)}$, $a=\{1,2\}$ and $x=\{A,B\}$, as 
\begin{align}
&U_a^{(x)}=e^{i\lambda}P_{1,2}P_{2,3}P_{3,4}U_{1,2}^{-1}\pi_{1,2}^{+}U_{1,3}^{-1}\pi_{1,2}^{-}
U_{2,3}^{-1}\pi_{3,4}^{+}\pi_{2,3}^{+}U_{1,4}^{-1}\pi_{2,3}^{-}U_{2,4}^{-1}\pi_{3,4}^{-}U_{3,4}^{-1} \, .
\label{eq333}
\end{align}
The phase global $e^{i\lambda}$  can be reabsorbed. The parameters of the matrices involved in the latter equation are listed in the Tables \ref{tab4} and \ref{table6}.
\begin{table}
\centering
\begin{tabular}{lllll}
\hline
$\mu,\nu$ & $\frac{\theta}{2}$ & $\beta$ & $\alpha$ & $\lambda$ \\ \hline
3,4                              &                        $-\frac{\pi}{4}$                            &                 $\frac{\pi}{2}$      &                       &       \\ 
2,4                    &                                 arctan($\frac{1}{\sqrt{2}}$)                             &        0          &                       &       \\
1,4                                   &   arctan($\frac{1}{\sqrt{3}}$)                                                   &                 $\frac{\pi}{2}$        &                       &       \\ 
2,3                                   &    0.91174                                                &       -0.46365     &                       &       \\ 
1,3                                   &            arctan(-$\frac{1}{\sqrt{2}}$)                                              &              $-\frac{\pi}{4}$          &                       &       \\ 
1,2                                   &        $\frac{\pi}{4}$                                              &             $\frac{\pi}{4}$           &                       &       \\
3,4                               &                                                    &                      &      $\frac{\pi}{8}$           &       \\ 
2,3                                   &                                             &                   &     -        &       \\ 
1,2                                   &                                                    &                      &       $\frac{3\pi}{8}$           &       \\ 
$\lambda$                               &                                                    &                      &                       &   $\frac{3\pi}{8}$   \\ \hline
\end{tabular}
\hspace{1cm}
\begin{tabular}{lllll}
\hline
$\mu,\nu$ & $\frac{\theta}{2}$ & $\beta$ & $\alpha$ & $\lambda$ \\ \hline
3,4                              &                        $-\frac{\pi}{4}$                            &                 $\frac{\pi}{4}$          &                       &       \\ 
2,4                    &                                 arctan(-$\frac{1}{\sqrt{2}}$)                             &           $\frac{\pi}{2}$          &                       &       \\ 
1,4                                   &   arctan($\frac{1}{\sqrt{3}}$)                                                   &                $-\frac{\pi}{4}$         &                       &       \\ 
2,3                                   &    0.91174                                                &             -1.249         &                       &       \\ 
1,3                                   &            arctan($\frac{1}{\sqrt{2}}$)                                              &              $\frac{\pi}{4}$          &                       &       \\ 
1,2                                   &        $\frac{\pi}{4}$                                              &           0         &                       &       \\ 
3,4                               &                                                    &                      &         $\frac{\pi}{2}$              &       \\
2,3                                   &                                             &                   &       $\frac{\pi}{2}$            &       \\ 
1,2                                   &                                                    &                      &        $ \frac{3\pi}{4} $             &       \\ 
$\lambda$                                &                                                    &                      &                       &   $\frac{3\pi}{4}$    \\ \hline
\end{tabular}
\caption{Parameters for the unitaries $U_{1}^{(A)}$ and $U_2^{(A)}$ involved in Eq. \eqref{eq333}.}
\label{tab4}
\end{table}
\begin{table}
\centering
\begin{tabular}{lllll}
\hline
$\mu,\nu$ & $\frac{\theta}{2}$ & $\beta$ & $\alpha$ & $\lambda$ \\ \hline
3,4                              &                        $\frac{\pi}{4}$                            &                 $\frac{3\pi}{8}$          &                       &       \\ 
2,4                    &                                 arctan($\frac{1}{\sqrt{2}}$)                             &           $\frac{\pi}{4}$          &                       &       \\ 
1,4                                   &   arctan($-\frac{1}{\sqrt{3}}$)                                                   &                $\frac{\pi}{8}$         &                       &       \\ 
2,3                                   &  0.91174                                               &        0.022583     &                       &       \\ 
1,3                                   &          arctan(-$\frac{1}{\sqrt{2}}$)                                              &        0            &                       &       \\ 
1,2                                   &        $\frac{\pi}{4}$                                              &           $-\frac{3\pi}{8}$          &                       &       \\ 
3,4                               &                                                    &                      &         $\frac{\pi}{8}$                &       \\ 
2,3                                   &                                             &                   &      $\frac{3\pi}{8}$                &       \\ 
1,2                                   &                                                    &                      &         $\frac{7\pi}{4} $             &       \\ 
$\lambda$                                &                                                    &                      &                       &   $-\frac{\pi}{4}$    \\ \hline
\end{tabular}
\hspace{1cm}
\begin{tabular}{lllll}
\hline
$\mu,\nu$ & $\frac{\theta}{2}$ & $\beta$ & $\alpha$ & $\lambda$ \\ \hline
3,4                              &                        $-\frac{\pi}{4}$                            &                 $-\frac{3\pi}{8}$          &                       &       \\ 
2,4                    &                                 arctan($\frac{1}{\sqrt{2}}$)                             &           $\frac{\pi}{4}$          &                       &       \\ 
1,4                                   &   arctan($\frac{1}{\sqrt{3}}$)                                                   &                $-\frac{\pi}{8}$         &                       &       \\ 
2,3                                   &   -2.2299                                             &       0.27258    &                       &       \\ 
1,3                                   &                  2.5261                                 &            $-\frac{\pi}{2}$        &                       &       \\ 
1,2                                   &        $\frac{\pi}{4}$                                              &           $-\frac{\pi}{8}$          &                       &       \\ 
3,4                               &                                                    &                      &         $\frac{11\pi}{16}$              &       \\ 
2,3                                   &                                             &                   &       $\frac{7\pi}{4}$            &       \\ 
1,2                                   &                                                    &                      &         $\frac{23\pi}{16} $             &       \\ 
$\lambda$                                &                                                    &                      &                       &   $\frac{7\pi}{16}$    \\ \hline
\end{tabular}
\caption{Parameters for the unitaries $U_{1}^{(B)}$ and $U_2^{(B)}$ involved in Eq. \eqref{eq333}.}
\label{table6}
\end{table}To reduce the evolution time of the system and consequently the effect of decoherence on it, the electromagnetic pulses  are implemented in parallel when possible. In particular, $P_{1,2}$ and $P_{3,4}$ are performed in parallel when implementing $U_{1}^{(A)}$, while $P_{3,4}$ and $U_{1,2}$ are performed simultaneously when implementing $U_{1}^{(B)}$, $U_{2}^{(A)}$ and $U_{2}^{(B)}$.\\ 
To verify the violation of the CGLMP inequality, we proceed as follows.
We consider simultaneous measurements of the two qudits.
We denote as $\rho_{A1,B1}$ the density matrix corresponding to the state obtained from the (previously-obtained approximation of the) maximally entangled state $\ket{\psi}$ in Eq. \eqref{starg32} the transformation $U_{1}^{(A)}$ (involving 14 gates) in Eq. \eqref{eq9supp} is applied on the first qubit, before the measurement, while  $U_{1}^{(B)}$ (consisting of 15 gates) is applied to the second qubit. In the same way, we also define the other density matrix $\rho_{A1,B2}$, $\rho_{A2,B1}$ and $\rho_{A2,B2}$: the fidelities between the four states resulting from the  procedure described above and their ideal counterparts are reported in Table \ref{fid}.\\
\begin{table}[!ht]
\centering
\begin{tabular}{llllll}
\hline

Amp & State & No decoherence & $T_2=30\mu s$ &$T_2=10\mu s$ &$T_2=5\mu s$ \\ \hline

0.004 & $\rho_{A_1,B_1}$ & 0.8157 & 0.7918 & 0.7473 &0.6873\\ 

0.004 & $\rho_{A_1,B_2}$ &0.8908  &  0.8656 & 0.8183 &  0.7546 \\ 

0.004 & $\rho_{A_2,B_1}$ & 0.6967 & 0.6764 & 0.6384 & 0.5872 \\ 

0.004 & $\rho_{A_2,B_2}$ &  0.6907 & 0.6700 & 0.6315 & 0.5798 \\ 

0.003 & $\rho_{A_1,B_1}$ & 0.8586  & 0.8297 & 0.7761 & 0.7051\\ 

0.003 & $\rho_{A_1,B_2}$&  0.9118 & 0.8822 & 0.8273 & 0.7542 \\ 

0.003 & $\rho_{A_2,B_1}$& 0.7586 & 0.7325  & 0.6842 &  0.6199 \\ 

0.003 & $\rho_{A_2,B_2}$ & 0.6884 & 0.6652 & 0.6223 & 0.5655 \\ 

0.002 & $\rho_{A_1,B_1}$ & 0.9187 & 0.8799 & 0.8092 & 0.7175 \\ 

0.002 & $\rho_{A_1,B_2}$ & 0.9304 & 0.8926 & 0.8234 & 0.7335 \\ 

0.002 & $\rho_{A_2,B_1}$& 0.9036 & 0.8652 & 0.7948 & 0.7034 \\ 

0.002 & $\rho_{A_2,B_2}$ & 0.8480 & 0.8129 & 0.7488 & 0.6656 \\ 

0.001 & $\rho_{A_1,B_1}$ & 0.9134  & 0.8528 & 0.7475 & 0.6214 \\ 

0.001 & $\rho_{A_1,B_2}$& 0.9183 &0.8597 & 0.7574 & 0.6339\\ 

0.001 & $\rho_{A_2,B_1}$ & 0.9389 &0.8755 & 0.7649 & 0.6319 \\ 

0.001 &$\rho_{A_2,B_2}$ & 0.9206 & 0.8615 & 0.7584 &0.6341\\ 

0.0009 & $\rho_{A_1,B_1}$& 0.9159 &0.8528 & 0.7473 &0.6061\\ 

0.0009 & $\rho_{A_1,B_2}$& 0.9160 & 0.8530 & 0.7441 & 0.6148 \\

0.0009 &$\rho_{A_2,B_1}$& 0.9483 & 0.8791 & 0.7598 & 0.6187 \\ 

0.0009 & $\rho_{A_2,B_2}$  & 0.9158 & 0.8527 & 0.7375 & 0.6140 \\ \hline

\end{tabular}
\caption{Fidelities between the density matrices of the states $\rho_{A_i B_j}$ prepared as explained in the text and their ideal counterparts.}
\label{fid}
\end{table} On the so-defined four states, local $S_z$ measurements were simulated by extracting the diagonal elements of the density matrix for each state, since the density matrix itself is taken decomposed on the eigenstates of $S_z$. The results are then combined to evaluate the functional $I$ defined in the main text.\\ Each of these operations involves a sequence of gates and takes a specific time to be implemented, as reported in Table \ref{tab9}, where the states have been organized based on the amplitude $B_1$ of the electromagnetic pulses of the form in Eq. \eqref{pulse} applied to obtain the unitaries in Eq. \eqref{eq9supp}. The findings are summarized in Table \ref{table10}. Additionally, Table \ref{tab11a} describes the outcomes when a further, non constant optimization of the same amplitudes is considered, allowing to achieve higher fidelities with the exact target states.\\
\begin{table}[h!]
\centering
\begin{tabular}{llllll}
\hline
State &  40G & 30G & 20G & 10G & 9G \\ \hline
$\rho_{A_1,B_1}$  & 82.23 & 109.64  & 164.45  & 328.91 & 365.45 \\ 
$\rho_{A_1,B_2}$  & 66.82 & 102.43  & 153.65 & 307.30 & 341.44 \\ 
$\rho_{A_2,B_1}$& 82.23 &  109.64 & 164.45 & 328.91  & 365.45  \\
$\rho_{A_2,B_2}$  & 66.82 & 102.43  & 153.65 & 307.30 & 341.44  \\ \hline
\end{tabular}
\caption{Simulation time (in ns) for amplitudes $B_1$ of the pulses in the range $[9-40]$ G.  The reported times must be added to the 127.84 ns required to achieve the maximally entangled state}
\label{tab9}
\end{table}
 \begin{table}[h!]
\centering
\begin{tabular}{lllll}
\hline
$B_1$(G) & No decoherence & $T_2=30\mu s$ & $T_2=10\mu s$ & $T_2=5\mu s$\\ \hline
40 & 2.1687 & 2.0910 & 2.0173 & 1.7544  \\
30 & 2.5129 &  2.4197 & 2.2471& 2.0191  \\
20 & 2.6443 & 2.5319 &  2.3259  & 2.0580 \\ 
10 & 2.7678 & 2.6103 & 2.3301 & 1.9806 \\
9 & 2.7425 & 2.5755 & 2.2027 & 1.9245 \\ \hline
\end{tabular}
\caption{CGLMP inequality violation. The reported values of the functional $I$ for varying decoherence conditions ($T_2=5,10,30,\infty$)  concern the situation when all the four states $\rho_{A1,B1}$, $\rho_{A1,B2}$, $\rho_{A2,B1}$ and $\rho_{A2,B2}$ are obtained with the same amplitude $B_1$ for all the transformations In Eq. \eqref{eq333}. Values of $B_1$ ranging from 9 G to 40 G have been considered. The inequality is violated when $I \geq 2$; the theoretical maximum value is $I=2.89624$.}
\label{table10}
\end{table}
\begin{table}[h!]
\centering
\begin{tabular}{ll}
\hline
\multicolumn{1}{c}{$B_1$ (G)} & \multicolumn{1}{c}{No decoherence} \\ \hline
20 [$\rho_{A_1,B_1/A_1,B_2}$] - 10 [$\rho_{A_2,B_1/A_2,B_2}$] & 2.7697 \\
20 [$\rho_{A_1,B_1/A_1,B_2}$] - 9 [$\rho_{A_2,B_1}$] - 10 [$\rho_{A_2,B_2}$] & 2.8030 \\
10 & 2.7678 \\
20 [$\rho_{A_1,B_1/A_1,B_2}$] - 9 [$\rho_{A_2,B_1/A_2,B_2}$] & 2.7573 \\ \hline
\end{tabular}
\vspace{0.3cm}  
\begin{tabular}{llll}
\hline
$T_2=30\mu s$ & $T_2=10\mu s$ & $T_2=5\mu s$ \\ \hline
2.6339 & 2.3889 & 2.1648 \\
2.6629 & 2.4107 & 2.0919 \\
2.6103 & 2.3301 & 1.9806 \\
2.6173 & 2.4317 & 2.0488 \\ \hline
\end{tabular}

\caption{The reported values of the functional $I$ for varying decoherence conditions (\( T_2 = 5, 10, 30, \infty \))  correspond to the situation where different amplitude values \( B_1 \) are used to implement the transformations in Eq.~\eqref{eq333}, leading to each of the four states \( \rho_{A1,B1} \), \( \rho_{A1,B2} \), \( \rho_{A2,B1} \), and \( \rho_{A2,B2} \). Specifically, different combinations of \( B_1 \) values ranging from 9 G to 20 G have been considered for each transformation.}
\label{tab11a}
\end{table}

\vspace{5cm}

\newpage
\bibliographystyle{apsrev4-2} 
\bibliography{biblio}

\end{document}